\documentclass[12pt]{article}

\usepackage{epsfig}

\usepackage{amsmath}
\usepackage{amssymb}
\usepackage{euscript}

\newcommand{\Peu}{\EuScript{P}}

\newcommand{\Deu}{\EuScript{D}}
\newcommand{\Reu}{\EuScript{R}}

\newcommand{\from}{\leftarrow}

\newcommand{\veps}{\varepsilon}

\begin{document}

\begin{titlepage}


\begin{flushright}
\bf IFJPAN-V-04-07\\
\bf CERN-PH-TH/2005-066
\end{flushright}

\vspace{1mm}
\begin{center}
  {\Large\bf%
    Solving constrained Markovian evolution in QCD\\
    with the help of the non-Markovian Monte Carlo$^{\star}$
}
\end{center}
\vspace{3mm}

\begin{center}
{\bf S. Jadach}
{\em and}
{\bf M. Skrzypek} \\

\vspace{1mm}
{\em Institute of Nuclear Physics, Academy of Sciences,\\
  ul. Radzikowskiego 152, 31-342 Cracow, Poland,}\\
and\\
{\em CERN Department of Physics, Theory Division\\
CH-1211 Geneva 23, Switzerland}
\end{center}

\vspace{5mm}
\begin{abstract}
We present the constrained Monte Carlo (CMC) algorithm
for the QCD evolution.
The constraint resides in that the total longitudinal
energy of the emissions in the MC and in the underlying QCD evolution is 
predefined (constrained).
This CMC implements exactly the full DGLAP evolution 
of the parton distributions in the hadron with respect
to the logarithm of the energy scale.
The algorithm of the CMC is referred to as the {\em non-Markovian} type.
The non-Markovian MC algorithm is defined as the one in which the 
multiplicity of emissions 
is chosen randomly as the first variable and
not the last one, as in the Markovian MC algorithms.
The former case resembles that of the fixed-order matrix element calculations.
The CMC algorithm can serve as an alternative to the so-called backward
evolution Markovian algorithm of Sj\"ostrand,
which is used for modelling the initial-state parton shower in 
modern QCD MC event generators.
We test practical feasibility and efficiency of our CMC implementation
in a series of numerical exercises,
comparing its results 
with those from other MC and non-MC programs,
in a wide range of $Q$ and $x$, down to the 0.1\% precision level.
In particular, satisfactory numerical agreement is found with
the results of the Markovian MC program of our own and
the other non-MC program.
The efficiency of the new constrained MC is found to be quite good.
\end{abstract}

\vspace{2mm}
\begin{center}
\em To appear in Computer Physics Communications
\end{center}

\vspace{5mm}
\begin{flushleft}
{\bf IFJPAN-V-04-07\\
\bf CERN-PH-TH/2005-066
}
\end{flushleft}

\vspace{3mm}
\footnoterule
\noindent
{\footnotesize
$^{\star}$This work is partly supported by the EU grant MTKD-CT-2004-510126
 in partnership with the CERN Physics Department and by the Polish Ministry
 of Scientific Research and Information Technology grant No 620/E-77/6.PR
 UE/DIE 188/2005-2008.
}

\end{titlepage}

\section{Introduction}
It is commonly known that the evolution equations of 
the quark and gluon distributions in the hadron,
derived in QCD by using the renormalization group or diagrammatic
techniques~\cite{DGLAP},
can be often interpreted probabilistically as a Markovian process;
see for instance the review of ref.~\cite{stirling-book}.

The Markovian type Monte Carlo (MC) algorithm%
\footnote{In the literature, see \cite{Kampen}, the Markovian process
  is usually understood as an infinite walk in a parameter space,
  in which the consecutive steps, indexed by means of 
  the time variable, obey the rule that every single
  step forward is independent of the past history of the walk.},
implementing the QCD or QED evolution equations,
is the basic ingredient in all parton-shower-type MCs.
An unconstrained forward Markovian MC,
with the evolution kernels from perturbative QCD/QED,
can only be used for the final-state radiation (FSR).
It is dramatically inefficient for the initial-state radiation (ISR),
because the hard process selects certain values of the parton energy
and the type of the proton.
The MC algorithm that allows us to fix (predefine)
the energy and the type of the parton exiting the
parton shower, entering the hard process, we shall call the 
{\em constrained MC algorithm}
or simply the {\em constrained MC} (CMC).

For the ISR parton shower an elegant example of the constrained MC
is the {\em backward Markovian evolution MC} algorithm of 
Sj\"ostrand~\cite{Sjostrand:1985xi}%
\footnote{The backward Markovian algorithm is also very well described in  ref.~\cite{Marchesini:1988cf}.},
which is a widely adopted effective solution in all popular MC event
generators, notably in PYTHIA~\cite{Sjostrand:2000wi} and HERWIG~\cite{Corcella:2000bw}.
However, the backward Markovian evolution algorithm is a kind of {\em work-around},
because it {\em does not solve}, strictly speaking,  the QCD evolution
equations.
It merely exploits their solutions coming from an {\em external},
typically non-MC, program.

The problem that we are posing and solving in the present work
is the following:
{\em Is it possible to invent an efficient constrained MC algorithm, 
not necessarily of the Markovian type, which
solves internally the full QCD DGLAP evolution equations on its own,
without relying on the external non-MC solutions?}
Since this is a highly non-trivial technical problem
in the area of the MC techniques, it is necessary to clearly state
the motivations that have led us to posing it and investing quite
some effort in finding at least one satisfactory solution.
The most important motivation is that we hope to gain
more power in the modelling of the ISR parton showers --
this can be potentially profitable for the better integration of the
complete next-to-leading-logarithmic (NLL) corrections in the parton
shower MCs. 
We also hope for an easier MC modelling of the
{\em unintegrated parton distributions} $D_k(Q,p_T,x)$
and MC modelling of the CCFM-type evolution~\cite{CCFM} in QCD. 

Let us define more precisely the terminology to be used in this work,
since it varies in the literature quite a lot.
By {\it Markovian MC algorithm} we understand
an algorithm in which the number of emissions 
(determining the dimension of the phase-space integral)
is generated as {\em  the last} variable or one of the last ones%
\footnote{Our definition of the Markovian MC
  implies that the corresponding Markovian process is terminated by means of 
  some
  stopping rule, for example by limiting the process time or other parameter.}.
On the other hand we shall call a {\it non-Markovian MC algorithm} 
the one in which the number of emissions
(the dimension of the integral),
is generated randomly as one of {\em the first} variables.

The name of  ``constrained MC'' can also be associated with a MC in which
the distribution to be generated is restricted to a less-dimensional
hyperspace by inserting the $\delta(F(x_1,...,x_n))$ function,
where $F(x_1,...,x_n)=0$ defines the constraint.
The CMC algorithm should efficiently generate points within this subspace.
It is widely known that it is usually much more complicated 
to generate  the distribution on such a hyperspace that in the entire space,
even if the original unconstrained distribution is simple, for example
it is the product of many simple distributions.
The energy-momentum conserving $\delta^{(4)}(P-\sum p_i)$ 
is a well known example of such a constraint.

In our case the constraint is given by $\delta(x-\prod z_i)$,
where $x$ is predefined 
(the outermost integration variable, generated in a CMC as the first one)
and $z_i$ are arguments of the DGLAP evolution kernels.
We shall describe an example of the CMC algorithm which 
fulfills such a constraint,
measure its efficiency and make a numerical test of 
its correctness, by comparing its results 
with those of the traditional unconstrained Markovian MC, 
and other non-MC programs.

The outline of the paper is the following:
in section 2 we introduce the basic notation and rederive an iterative solution
of the DGLAP evolution equations.
Since the full solution is algebraically involved,
we discuss in section 3 our CMC solution for the simpler case 
of pure brems\-strahlung out of quark or gluon.
Section 4 describes the hierarchical reorganization of the iterative
solution, which is necessary for the full scale solution with
arbitrary number of gluon--quark transitions. This complete solution,
along with numerical tests, is presented in section 5.
Summary and outlook concludes the paper.
For the sake of completeness, a small appendix lists the standard 
LL QCD kernels.

This paper describes the main part of a wider effort on the MC modelling
of the QCD evolution equations.
In ref.~\cite{raport04-06} the reader may find an alternative CMC
algorithm for the QCD evolution, which looks slightly inferior to the one
presented here and is implemented numerically only for pure bremsstrahlung.
References \cite{Jadach:2003bu} and  \cite{Golec-Biernat:2006xw}
are devoted to a precision MC modelling 
of the LL and NLL DGLAP evolution equations using the Markovian (unconstrained)
class of algorithms.
In this context it is worth to remind the reader that
the consistent integration of the NLL perturbative corrections
at the fully exclusive level in the MC event generator of the parton shower
type is the challenging problem both theoretically and practically.
For the recent efforts in this direction
see refs.~\cite{Frixione:2002ik,Nason:2004rx}, for example.
The algebraic proof of certain important identity used in this work
is published separately in ref.~\cite{raport04-09}.

\section{Iterative solution of the QCD evolution equations}
\label{sec:itersolution}
Let us rederive the {\em iterative solution} 
of the QCD evolution equations. 
The starting point is the DGLAP~\cite{DGLAP} set of evolution equations:
\begin{equation}
\begin{split}
  \label{eq:Evolu}
  \frac{\partial}{\partial t} D_k(t,x)
  &= \sum_j \int\limits_x^1 \frac{d z}{z} P_{kj}(z) 
    \frac{\alpha_S(t)}{\pi} D_j\Big(t,\frac{x}{z} \Big)
= \sum_j {\Peu}_{kj}(t,\cdot)\otimes D_j(t,\cdot),
\end{split}
\end{equation}
where
\begin{equation}
f(\cdot){\otimes} g(\cdot)(x) 
    \equiv \int dx_1 dx_2 \delta(x-x_1 x_2)f(x_1)g(x_2),
\end{equation}
and 
${\Peu}_{kj}(t,z)\equiv \frac{\alpha_S(t)}{\pi}  P_{kj}(z)$.
Indices $i$ and $k=G,q,\bar q$ denote gluon or quark --
the evolution { time} is $t=\ln(Q)$.
The differential evolution equation can be turned into the 
following integral equation:
\begin{equation}
\begin{split}
 & e^{\Phi_k(t,t_0)}   D_k(t,x)
    = D_k(t_0,x)
   +\int\limits_{t_0}^t dt_1  e^{\Phi_k(t_1,t_0)}
    \sum_j {\Peu^\Theta_{kj}(t_1,\cdot)}
    \otimes D_j(t_1,\cdot)(x),
\end{split}
\end{equation}
where $\veps$ is  the IR regulator in the kernels,
\begin{eqnarray}
  {\Peu_{kj}(t,z)}&=&-\Peu^{\delta}_{kk}(t,\veps) \delta_{kj}\delta(1-z)
                 +\Peu^{\Theta}_{kj}(t,z),
\\
  \Peu^{\Theta}_{kj}(t,z)&=&\Peu_{kj}(t,z)\Theta(1-z-\veps),
\end{eqnarray}
and the Sudakov form factor is given by
\begin{equation}
    {\Phi_{k}(t,t_0)} = \int_{t_0}^t  dt'\;
       \Peu^\delta_{kk}(t',\veps).
\end{equation}
The complete set of LL kernels $\Peu_{kj}$ is collected in an appendix.

The multiple iteration of the above integral equation leads us to 
an {\it iterative solution} of the evolution equations, given by
the following series of integrals, 
ready for the evaluation with the help of the standard MC methods:
\begin{equation}
\label{eq:iter1}
\begin{split}
 &D_k(t,x) = e^{-\Phi_k(t,t_0)} D_k(t_0,x)
  +\sum_{n=1}^\infty \;
   \sum_{k_0...k_{n-1}}
      \bigg[ \prod_{i=1}^n \int\limits_{t_0}^t dt_i\;
      \Theta(t_i-t_{i-1})  \int\limits_0^1 dz_i\bigg]
\\&
      e^{-\Phi_k(t,t_n)}
      \int\limits_0^1 dx_0\;
      \bigg[\prod_{i=1}^n 
           \Peu_{k_ik_{i-1}}^\Theta (t_i,z_i) 
            e^{-\Phi_{k_{i-1}}(t_i,t_{i-1})} \bigg]
     D_{k_0}(t_0,x_0) \delta\bigg(x- x_0\prod_{i=1}^n z_i \bigg),
  \end{split}
\end{equation}
where $k_n\equiv k$.
In ref.~\cite{Jadach:2003bu} the above equation
was solved with the three-digit precision using the Markovian MC method%
\footnote{The energy sum rules $\sum_l \int dz\; z\Peu_{lk}(z)=0$ 
  are instrumental in this MC method and in fact the equation for
$xD_k(x)$ rather than $D_k(x)$ is solved there.}.
In this work the above solution 
is {\em the starting point} for constructing the CMC algorithm.

There is still one standard technical point:
the one-loop dependence of the strong coupling
$\alpha_S(t)= 2\pi/( \beta_0 (t-\ln\Lambda_0))$
may destroy the efficiency of the MC algorithm, unless
it is conveniently compensated by means of 
the mapping $t_i\to \tau_i=\ln(t_i-\ln\Lambda_0)$.
Once it is done, the iterative solution transforms into the following form:
\begin{equation}
  \label{eq:Iter5}
  \begin{split}
  D_k&(\tau,x) = e^{-\Phi_k(\tau,\tau_0)} D_k(\tau_0,x)
  +\sum_{n=1}^\infty \;
  \int_0^1 dx_0\;
   \sum_{k_0...k_{n-1}}
   \bigg[ \prod_{i=1}^n  \int_{\tau_0}^\tau d\tau_i\; 
     \Theta(\tau_i-\tau_{i-1}) \int_0^1 dz_i\bigg]
\\&~~\times
      e^{-\Phi_k(\tau,\tau_n)}
      \bigg[ \prod_{i=1}^n 
           \Peu_{k_ik_{i-1}}^\Theta (\tau_i,z_i) 
                 e^{-\Phi_{k_{i-1}}(\tau_i,\tau_{i-1})} \bigg]
      D_{k_0}(\tau_0,x_0) \delta\bigg(x- x_0\prod_{i=1}^n z_i \bigg),
  \end{split}
\end{equation}
where $k\equiv k_n$. 
The kernel $\Peu$ and form factor $\Phi$ are from now on redefined as follows:
\begin{equation}
\begin{split}
  \Peu_{k_ik_{i-1}} (\tau_i,z_i)
  &=\frac{\alpha_S(t)}{\pi}\;\frac{\partial t}{\partial \tau}\; 
                       P_{k_ik_{i-1}} (z_i)
  =\frac{2}{\beta_0}\; P_{k_ik_{i-1}} (z_i),
\\
    {\Phi_{k}(\tau,\tau_0)} &= \int_{\tau_0}^\tau  d\tau'\;
       \Peu^\delta_{kk}(\veps)
        = (\tau-\tau_0)\Peu^\delta_{kk}(\veps).
\end{split}
\end{equation}
In the present LL case, $\Peu$ becomes completely independent of $\tau_i$.
See also refs.~\cite{raport04-06,Golec-Biernat:2006xw}
for more discussion on the optimal choice of the evolution time.

In the following we shall often employ the short-hand notation:
\begin{equation}
\begin{split}
&\theta_{x>y}\equiv\theta(x-y),\;\;
   \hbox{where}\;\; \theta_{x>y}=1\;\; \hbox{for}\;\; x>y,\;\; 
   \hbox{otherwise}\;\;   \theta_{x>y}=0,
\\
&\delta_{x=y}\;\; \hbox{instead of}\;\; \delta(x-y),
\end{split}
\end{equation}
so as to keep formulas more compact.

Throughout this work we concentrate on the LL case.
However, our CMC algorithm solves most of the problems on the way
to the CMC solution for the NLL DGLAP.
The exact Monte Carlo solution of the NLL DGLAP evolution equations
in the framework of the unconstrained Markovian approach is presented
in a separate work (ref.~\cite{Golec-Biernat:2006xw}). 
It may serve as a numerical benchmark for the future work on the NLL CMC.

\section{CMC for pure bremsstrahlung only}
\label{sec:cmc-brems}
Before we unfold all the details of our construction of the CMC 
algorithm for the full DGLAP equations,
let us first describe it
for the simpler case of the pure QCD brems\-strahlung.
This will help the reader to follow all algebraic technicalities
of the full CMC solution.
It will also provide a building block for the full CMC solution.

The starting point is the iterative solution of the QCD evolution
equations of eq.~(\ref{eq:Iter5}) truncated 
to the multiple gluon brems\-strahlung emitted from the 
parton $k=G,q,\bar{q}$,
with the starting distribution $D_k(\tau_0,x_0)=\delta(x_0-1)$:
\begin{equation}
\label{eq:iterative}
  \begin{split}
 &x\Deu_{kk}(\tau,\tau_0;x) = 
  e^{-\Phi_{k}(\tau,\tau_0)} \bigg\{ \delta_{x=1}+
  \sum_{n=1}^\infty \frac{1}{n!} \;
   \prod_{i=1}^n 
     \int\limits_{\tau_{i-1}}^\tau d\tau_i\;  
     \int\limits_0^1 dz_i\;z_i
	\Peu_{kk}^\Theta (\tau_i,z_i) 
      \delta_{x=\prod_{i=1}^n z_i} \bigg\}.
\\ 
  \end{split}
\end{equation}
For the purpose of the MC generation, we introduce simplified kernels:
\begin{equation}
\label{eq:truncation}
\begin{split}
  & zP^{\Theta}_{GG}(z) \to z\hat P^{\Theta}_{GG}(z)=
  zB_{GG}\Theta(1-\veps-z)
      \left(\frac{1}{1-z}+\frac{1}{z}\right)
   =  B_{GG}\frac{\Theta(1-\veps-z)}{1-z},
\\&
   zP^{\Theta}_{qq}(z) \to z\hat P^{\Theta}_{qq}(z)
   =  B_{qq}\frac{\Theta(1-\veps-z)}{1-z}.
\end{split}
\end{equation}
Note that in eq.~(\ref{eq:truncation}) we do not modify the virtual
parts of the kernels. We also {\em do not} assume any sum rules relating
$\Peu^\delta_{kk}$ to $\int z \Peu^\theta_{kk}$. This is so because we
view eq.~(\ref{eq:iterative}) as a part of a bigger framework (eq.\
(\ref{eq:Iter5}) for example) with both quarks and gluons, and we will
assume later on a {\em complete} sum rule 
$\Peu^\delta_{kk} =\sum_j\int z\Peu^\theta_{jk}$.

The above simplification of eq.~(\ref{eq:truncation})
will be countered by the MC weight $w_P$ defined below:
\begin{equation}
  \begin{split}
 &x\Deu_{kk}(\tau,\tau_0;x) = e^{-\Phi_{k}(\tau,\tau_0)} 
  \bigg\{ \delta_{x=1}+
  x^{\omega_k}
  \sum_{n=1}^\infty \frac{1}{n!} 
  b_k^n\;
   \prod_{i=1}^n 
     \int\limits_{\ln(\veps)}^{\ln(1-x)} dy_i\;  
     \delta_{x=\prod_{i=1}^n z_i(y_i)}
     \int\limits_{\tau_{i-1}}^{\tau} d\tau_i\; w_P
 \bigg\},
  \end{split}
\end{equation}
where $y_i=\ln(1-z_i)$ and 
\begin{equation}
\begin{split}
  & w_P= x^{-\omega_k}
  \prod_{j=1}^n \frac{P^{\Theta}_{kk}(z_j)}{\hat P^{\Theta}_{kk}(z_j)},
\\& \Phi_{k}(\tau,\tau_0)
      =(\tau-\tau_0) \left( b_{k} \ln\frac{1}{\veps} -a_k \right),
\\& b_k\equiv \frac{2}{\beta_0} B_{kk},\quad
    a_k\equiv \frac{2}{\beta_0} A_{kk}.
\end{split}
\end{equation}
The factor $x^{\omega_k}$ is introduced in order to obtain as uniform shape of
the MC weight distribution for all $x\in (0,1)$ as possible.
It will also help to correct for the fact 
that $P_{qq}(z_i)$ does not have a $1/z_i$ component --
in this case $\prod_i z_i =x $ can be pulled out of the integrand.
For the moment we assume $\omega_G=0$ and $\omega_q=1$, 
but obviously they are dummy parameters, which may influence
the MC efficiency but not the final MC distributions.

The energy constraint
\begin{equation}
  \begin{split}
   &x=\prod_{i=1}^n z_i(y_i)=
    \prod_{i=1}^n \big(1-\exp(y_i)\big)=F(y_1,y_2,\dots,y_n),
    \equiv F({\bf y})
    \quad \hbox{\rm or}
\\&
  \ln\frac{1}{x}=
    \sum_{j=1}^n f(y_j)=-\ln F({\bf y});
    \quad 
    f(y_j) =-\ln\big(1-\exp\big(y_j)\big)=-\ln z_j
  \end{split}
\end{equation}
provides also an integration limit $z(y_i)\geq x$, 
which translates into $y_i=\ln(1-z_i)\leq \ln(1-x) $.
Taking advantage of the symmetry of the integrand we can introduce ordering 
in the $y$-variables:
\begin{equation}
  \begin{split}
 &x\Deu_{kk}(\tau,\tau_0;x) = e^{-\Phi_{k}(\tau,\tau_0)} 
  \bigg\{ \delta_{x=1}+
\\&~~~+
  x^{\omega_k-1}\sum_{n=1}^\infty b_k^n\;
   \prod_{i=1}^n 
     \int\limits_{y_{\min}}^{y_{\max}} dy_i\;
     \theta_{y_i>y_{i-1}} 
     \delta\left(\ln\frac{1}{x}-\sum_j f(y_j)\right)
     \int\limits_{\tau_0}^{\tau} d\tau_i\; w_P
 \bigg\}.\;
  \end{split}
\end{equation}
where $y_0\equiv y_{\min}=\ln(\veps)$ and  $y_{\max}=\ln(1-x)$.

The function $f(y_i)$ is very steeply (exponentially) rising,
hence the constraint $x=\prod_{i=1}^n z_i(y_i)$ is effectively ``saturated'' 
by a single $z_j$
while the other ones are $z_i\simeq 1,\; i\neq j$.
In terms of $y$ variables, $y_j\simeq y_{\max}$,
while the other ones $y_i,\;\; i\neq j$, move freely 
within the $(y_{\min},y_{\max})$ interval.
In our case, because we ordered $y$-variables, it is the biggest 
$y_n \simeq y_{\max}$
that effectively takes responsibility for satisfying the constraint.

\begin{figure}[!h]
\centering
  {\epsfig{file=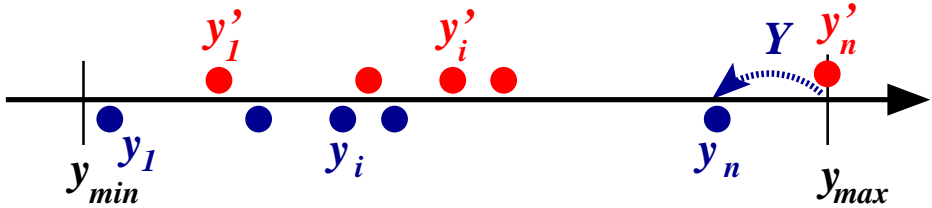, width=130mm}}
\caption{Graphical representation of the linear transformation 
         $y_i \to y_i'$}
\label{fig:shift}
\end{figure}

In the following we translate the above statements into rigorous mathematics.
The procedure of eliminating the energy constraint goes in three steps:
\\
{\em Step one.}
We introduce the new  integration variable $Y$.
Its introduction is immediately countered by the $\delta$-function:
\begin{equation}
  \begin{split}
 &x\Deu_{kk}(\tau,\tau_0;x) = e^{-\Phi_{k}(\tau,\tau_0)} 
  \bigg\{ \delta_{x=1}+
  x^{\omega_k-1}\sum_{n=1}^\infty b_k^n\;
\\&~~~ \times
   \int dY
   \prod_{i=1}^{n} 
     \int\limits_{y_{\min}}^{y_{\max}} dy_i\;
     \theta_{y_i>y_{i-1}}
     \delta(y_n+Y-y_{\max})
     \delta\left(\ln\frac{1}{x}-\sum_j f(y_j)\right)
     \int\limits_{\tau_0}^{\tau} d\tau_i\; w_P.
 \bigg\}.\;
  \end{split}
\end{equation}
{\em Step two.}
We perform the following simple linear transformation
\begin{equation}
  y_i=y'_i-Y.
\end{equation}
Note that $Y$ was ``adjusted'' from the very beginning
such that $y'_{n}=y_n+Y=y_{\max}$, 
see also a graphical illustration in fig.~\ref{fig:shift}.

The Jacobian of the above linear transformation is equal to 1. We obtain:
\begin{equation}
  \begin{split}
 &x\Deu_{kk}(\tau,\tau_0;x) = e^{-\Phi_{k}(\tau,\tau_0)} 
  \bigg\{ \delta_{x=1}+
  x^{\omega_k-1}\sum_{n=1}^\infty b_k^n\;
   \int dY
\\&~~~~~~~~\times
   \prod_{i=1}^{n} 
     \int\limits_{y_{\min}}^{y_{\max}} dy'_i\;
     \theta_{y'_i>y'_{i-1}}
     \delta(y'_n-y_{\max})
     \delta\left(\ln\frac{1}{x}-\sum_j f(y'_j-Y)\right)
     \int\limits_{\tau_0}^{\tau} d\tau_i\; w_P
 \bigg\},
  \end{split}
\end{equation}
where $y'_0\equiv y_{\min}+Y$,
and $\theta_{y_1>y_{\min}}=\theta_{y'_1>Y+y_{\min}}=\theta_{y'_1>y'_0}$
defines the IR lower boundary of the phase space.
\\
{\em Step three.}
The energy constraint $\delta(x-F)$ is eliminated
by means of the $Y$-integration:
\begin{equation}
  \begin{split}
 &x\Deu_{kk}(\tau,\tau_0;x) = e^{-\Phi_{k}(\tau,\tau_0)} 
  \bigg\{ \delta_{x=1}+
\\&~~~+
  x^{\omega_k-1}\sum_{n=1}^\infty b_k^n\;
   \prod_{i=1}^{n} 
     \int\limits_{y_{\min}}^{y_{\max}} dy'_i\;
     \theta_{y'_i>y'_{i-1}}
     \delta(y'_n-y_{\max})
     \frac{1}{|\partial_Y \ln F({\bf y}'-Y)|_{Y=Y_0}}
     \int\limits_{\tau_0}^{\tau} d\tau_i\; w_P
 \bigg\},
  \end{split}
\end{equation}
where $Y_0=Y_0(x,{\bf y}')$ is the solution of the transcendental equation
$x=F({\bf y}'-Y)$. 
One subtle point is that thanks to $y'_0=y_{\min}+Y_0(x,{\bf y}')$, and
$Y_0\geq 0$  
we may keep formally the same integration limits 
$y'_i\in (y_{\min},y_{\max})$ 
 as before.

Summarizing the above three steps:
we effectively traded complicated $\delta(x-F({\bf y}))$ 
into much simpler $\delta(y'_n-y_{\max})$.
We may also say that we projected, by means of the parallel shift%
\footnote{The projection of a similar type 
 (with dilatation instead of parallel shift)
 has been employed in the CMC already in refs.~\cite{VanHoveKittel},
while the idea 
 of saturating the constraint with a simpler one using single variable can
 be traced back to the MC program of 
 ref.~\cite{genrap} (and its unpublished versions)
 as well as to ref.~\cite{Jadach-yfs-mpi:1987}.}
(along the diagonal),
points from the hyperspace $y_n=y_{\max}$ into the hyperspace $x=F({\bf y})$.

Let us now have a closer look into formula for $\partial_Y F({\bf y}'-Y)$,
which is necessary for the MC weight and for the numerical solution of
the equation $x=F({\bf y}'-Y)$:
\begin{equation}
  \begin{split}
  &|\partial_Y \ln F({\bf y}'-Y)|_{Y=Y_0}
    =\left|\left(\sum_{k=1}^n\partial_Y f(y_k'-Y)\right)\right|_{Y=Y_0}
    =\bigg| \sum_{k=1}^n 
      \frac{\exp(y_k'-Y)}{z_k} 
      \bigg|_{Y=Y_0}
\\&
    = \sum_{k=1}^n 
      \frac{1-z_k}{z_k}\;
    = \sum_{k=1}^n |\partial_y \ln z(y)|_{z=z_k}
  \end{split}
\end{equation}

The additional transformation $y'_i=y_{\min}+r_i\Delta_y$ with
$\Delta_y=y_{\max}-y_{\min}$ and $0\leq r_i\leq 1$
as well as exporting part
of the integrand into MC weight brings the integral closer to the form
suited for the MC generation:
\begin{equation}
  \begin{split}
 &x\Deu_{kk}(\tau,\tau_0;x) = e^{-\Phi_{k}(\tau,\tau_0)} 
  \bigg\{ \delta_{x=1}+
 x^{\omega_k-1}\sum_{n=1}^\infty b_k^n \Delta^{n-1}_y
   \prod_{i=1}^{n} 
     \int\limits_{0}^{1} dr_i\;
     \theta_{r_i>r_{i-1}}
     \delta(r_n-1)
     \int\limits_{\tau_0}^{\tau} d\tau_i
     w_0^{\#}
 \bigg\},
\\& w_0^{\#}= w_P \frac{1}{|\partial_Y \ln F({\bf y}'-Y)|_{Y=Y_0}}\; 
    \theta_{y'_1-Y_0>y_{\min}},
  \quad r_0\equiv 0.
  \end{split}
\end{equation}
The average or maximum weight is improved if the weight $w_0^{\#}(z_1,...,z_n)$
is rescaled by the function $g(x)$
(which can be pulled out of the integral).
The weight distribution is conveniently stabilized if we take for $g(x)$
the biggest term in $|\partial_y \ln z(y)|_{z=z_k}$, with the largest $y_k$,
which we approximate as $z_k\simeq x$:
\begin{equation}
  \begin{split}
   &w^{\#}= w_P\; \frac{xg(x)}%
                 {|\partial_Y \ln F({\bf y}'-Y)|_{Y=Y_0}}\; 
                 \theta_{y'_1-Y_0>y_{\min}},
\quad
 g(x)=|\partial_y \ln z(y)|_{z=x}
   =\frac{1-x}{x}.
  \end{split}
\end{equation}

Finally we introduce the normalized variables 
$s_i=(\tau_i-\tau_0)/(\tau-\tau_0)$,
rescale $w^{\#}$ and symmetrize over $r_i$:
\begin{equation}
  \begin{split}
 &x\Deu_{kk}(\tau,\tau_0;x) = e^{-\Phi_{k}(\tau,\tau_0)} 
  \bigg\{ \delta_{x=1}+
\\&~~~~~~~~~~+
  \frac{x^{\omega_k-1}}{xg(x)}
  \sum_{n=1}^\infty 
  b_k \frac{\Delta^{n-1}}{(n-1)!}\;
   (\tau-\tau_0)
   \prod_{i=1}^{n} 
     \int\limits_{0}^{1} dr_i\;
     \frac{\delta(1- \max r_j)}{n}
     \int\limits_0^1 ds_i\; w^{\#}
 \bigg\},
\\& \Delta
    =b_k(\tau-\tau_0)\Delta_y
    =b_k(\tau-\tau_0)[\ln(1-x)-\ln\veps]
    =\Reu(1-x)-\Reu(\veps),\; 
\\& 
   \Reu(x)=b_k(\tau-\tau_0)\ln x.
  \end{split}
\end{equation}
Introducing explicitly a Poisson distribution
$P(n|\lambda)=e^{-\lambda}\lambda^n/n!$,
and remembering that 
$\Phi_k(\tau,\tau_0)= -(b_k\ln\veps +a_k)(\tau-\tau_0)$,
we obtain the following {\em master formula} on which the MC algorithm
is built:
\begin{equation}
\begin{split}
 &x\Deu_{kk}(\tau,\tau_0;x) =
  \sum_{n=0}^\infty 
  e^{(\tau-\tau_0)a_k}
  \bigg\{ 
  e^{\Reu(\veps)} 
  \delta_{n=0}\delta_{x=1}
+\delta_{n>0}\theta_{1-x>\veps}
 e^{\Reu(1-x)} 
  \frac{b_k x^{\omega_k-1} }{xg(x)}(\tau-\tau_0)
\\&~~~~~~~~~~\times
  P\Big(n-1 \Big| \Reu(1-x)-\Reu(\veps)\Big)
   \prod_{i=1}^{n} 
     \int\limits_{0}^{1} dr_i\;
     \frac{\delta(1- \max r_j)}{n}
     \int\limits_0^1 ds_i\; w^{\#}
 \bigg\},
\\&
  \tau_i = \tau_0 +s_i(\tau-\tau_0),\quad
  z_i = 1-e^{y_i} = 1-\exp(y_{\min} +r_i \Delta_y-Y_0).
\end{split}
\end{equation}

In our CMC algorithm for pure bremsstrahlung
one generates first $x$, then $n$, next $z_i$ (constructed from $y_i$) 
and finally 
$\tau_i$ (constructed from $s_i$).
The algorithm is clearly non-Markovian, as is seen from the fact that the
number of emissions is generated as the second variable, not as the last one.
The distribution of the variable $x$ is done
according to the following {\em primary distribution}:
\begin{equation}
\label{deuprim}
\begin{split}
 x\Deu'_{kk}(\tau,\tau_0;x) &= 
  e^{(\tau-\tau_0)a_k} \bigg\{ 
   \delta_{x=1}\; e^{\Reu(\veps)}
 + \theta_{1-x>\veps}
   e^{\Reu(1-x)}\;
  \frac{b_k(\tau-\tau_0)}{xg(x)}\;x^{\omega_k-1} \bigg\}
\\
&= 
  e^{(\tau-\tau_0)a_k} \bigg\{ 
  \delta_{x=1}\; e^{\Reu(\veps)}
  + \theta_{1-x>\veps} \;
   e^{\Reu(1-x)}
   | \partial_x\Reu(1-x)|\; x^{\omega_k-1} \bigg\},
\end{split}
\end{equation}
which is obtained
by means of neglecting $w^{\#}$ and performing all summations and integrations.
The MC weight $w^{\#}$ is, of course, restored later on.

It is quite convenient, in the construction of the MC program, that
in the above distribution we are able to extend artificially
the integration above $x=1-\veps$, by means of mapping
$x$ into the new variable $U=\exp( \Reu(1-x))$,
so that we can generate $x$ as if there was no $\delta_{x=1}$.
This resembles quite strongly the analogous trick for
the multiple photon emissions in the YFS-type MCs for QED~\cite{yfs2:1990}.
Let us work out the details, restoring the initial parton distribution
at $t=t_0$ and the hard process function $H(x)$:
\begin{equation}
\begin{split}
& \int\limits_{\epsilon_1}^{1} dx\;  H(x)\; D_{kk}(\tau,x)=
  \int\limits_{\epsilon_1}^{1} dx\;
  \int\limits_{0}^{1} dZ\; 
  \int\limits_{0}^{1} dx_0\; 
  \delta(x-x_0Z\;)
  H(x)\; \Deu'_{kk}(\tau,\tau_0;Z)\; D_k(\tau_0,x_0)\;
\\&
 =\int\limits_{\epsilon_1}^{1} dx H(x)
  \int\limits_{x}^{1} \frac{dZ}{Z}
     Z^{\omega_k-2} 
      e^{(\tau-\tau_0)a_k}
    \left\{
    \delta_{Z=1}\; e^{\Reu(\veps)}
    + \theta_{1-Z>\veps} 
    e^{\Reu(1-Z)}
   |\partial_Z \Reu(1-Z)|
    \right\}
  D_k\Big(\tau_0,\frac{x}{Z}\Big)
\\&
   =\int\limits_{\epsilon_1}^{1} dx\; H(x)
    \int\limits_0^{\exp(\Reu(1-x))}\!\!\! dU\;
    Z(U)^{\omega_k-3}  e^{(\tau-\tau_0)a_k}
    \left\{
    \delta_{U=0}\; e^{\Reu(\veps)}
    +\theta_{U\geq \exp(\Reu(\veps))} 
    \right\}
    D_k\Big(\tau_0,\frac{x}{Z(U)}\Big)
\\&~~
  =\int\limits_{\epsilon_1}^{1} \frac{dx}{x}\; H(x)\;
   \int\limits_0^{\exp(\Reu(1-x))}\!\!\! dU\; 
    Z(U)^{\omega_k-2} e^{(\tau-\tau_0)a_k}\;\;
    \frac{x}{Z(U)} D_k\Big(\tau_0,\frac{x}{Z(U)}\Big),
\\& U(Z)=e^{\Reu(1-Z)}=(1-Z)^{b_k(\tau-\tau_0)},
\\& Z(U)=1-\exp\big((b_k(\tau-\tau_0))^{-1}\ln U \big),
\end{split}
\end{equation}
and remembering that $1>Z(U)>1-\veps$ is mapped
exactly into one point at $U=0$,
reproducing the component $\sim\delta_{Z=1}$, i.e.
$\int^{\exp(\Reu(\veps))}_0 dU =\exp(\Reu(\veps))$.
Once $x$ is chosen, $n$ is generated according to a Poisson
distribution (shifted by 1) and then all variables $\tau_i$ and $z_i$
are generated. 

\begin{figure}[!h]
\centering
  {\epsfig{file=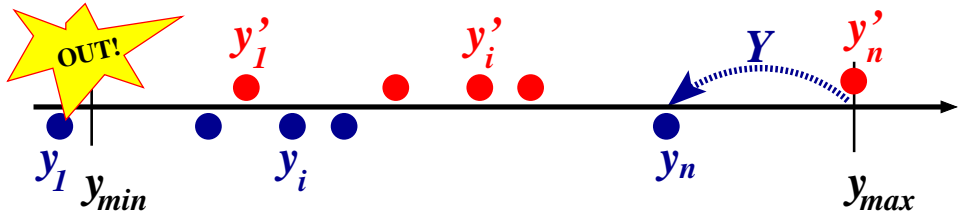,width=130mm}}
\caption{Graphical representation of the rescaling procedure $y_i\to y_i'$}
\label{fig:rescaling}
\end{figure}

The above collection of detailed formulas determines uniquely the whole CMC
algorithm for the pure bremsstrahlung case%
\footnote{The only element that is not described in fine detail is the
  method of solving the transcendental equation for the constraint $Y_0(y'_i)$.
  We use a variant of the standard method of the tangentials.
  It is in principle quite straightforward -- the only complication
  is that it must work for all values of $y'_i$.
  For certain values, because of $\partial_Y F \sim 0 $,
  attention has to be paid that the 
  number of the iterations is sufficient.}.
For the sake of completeness let us summarize point by point 
the complete CMC algorithm:
\begin{itemize}
\item The outmost integration variable generated as the first one
  is total $x$, the argument of the hard process cross section $H(x)$.
\item The second generated variable is $Z$, the total loss of energy
  due to multiple gluon brems\-strahlung,
  with the help of the mapping: 
  $U(Z)=e^{\Reu(1-Z)}=(1-Z)^{b_k(\tau-\tau_0)}$.
\item The generation of variables $x$ and $U$ (and of $k=k_0$, if necessary)
  is done with the help of the general-purpose MC tool {\tt FOAM}
  \cite{foam:1999, foam:2002}.
\item Knowing $Z(U)$, if $Z>1-\veps$, {\em the emission multiplicity $n$}
  is generated according to a Poisson distribution $P_{n-1}$ (non-Markovian!), 
  otherwise $Z=1$ and $n=0$.
\item Variables $s_i, i=1,2,...,n$ are generated uniformly and mapped 
  onto $\tau_i(s_i)$ and $t_i(\tau_i)$.
  They are ordered.
\item Not ordered variables $r_i\in (0,1)$ are generated, 
  such that one of them is set equal to%
  \footnote{We generate $n$ of them and rescale 
  such that the biggest is equal~1.}
  1; they are mapped into $y'_i(r_i)$.
\item The solution $Y=Y_0$ of the transcendental equation 
$\ln F({\bf y}'-Y)-\ln x =0$
  is found numerically
  (NB: the derivative $\partial_Y\ln F$ for the MC weight is 
   obtained as a byproduct).
\item With $Y_0$ at hand, all variables $z_i(y_i(y'_i)),\; i=1,2,...,n$ 
  are calculated.
\item In the case $y_1<y_{\min}$, see fig.~\ref{fig:rescaling},
  the MC weight $w^\#$ is zero and the algorithm stops.
\item Finally the MC weight $w^\#$ is calculated. Optionally
  the weighted MC event is transformed into an unweighted one 
  with the help of the standard rejection method.
\end{itemize}

In this way we completed the description of the CMC algorithm for the pure
brems\-strahlung case, which will be the essential building block for the
general-case CMC algorithm described in the following.

\subsection{Numerical test of the CMC for pure bremsstrahlung}
\label{sec:numtest0}
\begin{figure}[!ht]
  \centering
  {\epsfig{file=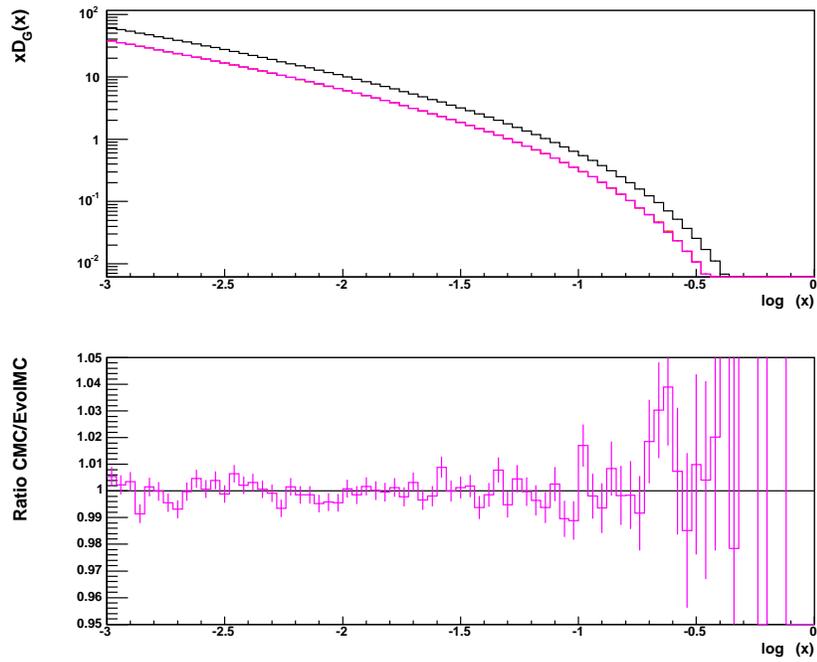,width=120mm}}
  \caption{\sf
    Comparison of CMC and EvolMC in the case of pure 
    bremsstrahlung out of gluon.
    }
  \label{fig:bremsCMC_glue}
\end{figure}
\begin{figure}[!ht]
  \centering
  {\epsfig{file=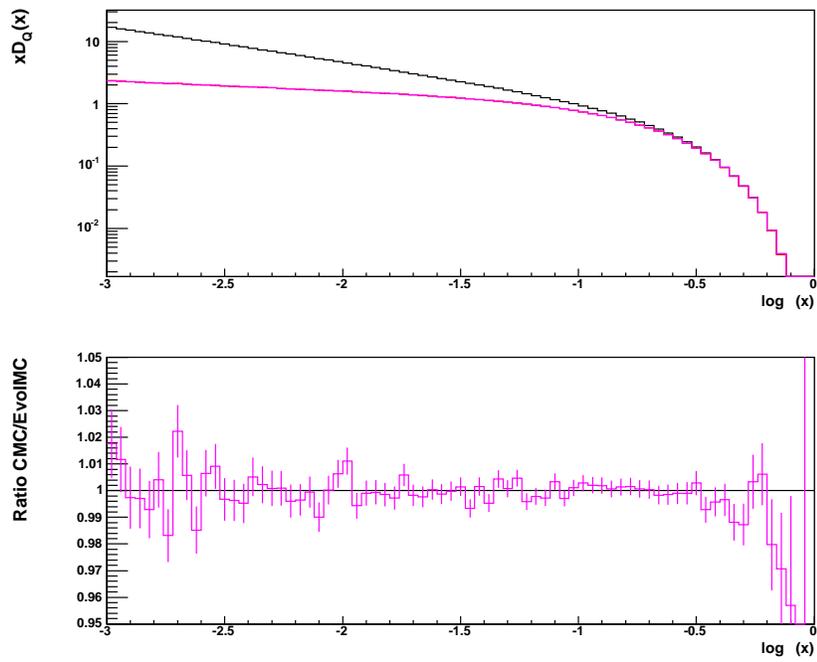,width=120mm}}
  \caption{\sf
    Comparison of CMC and EvolMC in the case of pure 
    bremsstrahlung out of quark
    }
  \label{fig:bremsCMC_quark}
\end{figure} 
The CMC for pure gluon bremsstrahlung was tested separately
for emission from gluon and quark lines
by comparing its results with those from the Markovian MC {\tt EvolMC}
of refs.~\cite{Jadach:2003bu} and \cite{Golec-Biernat:2006xw}.
As in ref.~\cite{Jadach:2003bu} results are for the LL DGLAP evolution
of the parton distribution in the proton 
from energy scale $Q=1$ GeV to $Q=1$ TeV.
We use the same starting parton distributions in proton at $Q=1$ GeV
as in ref.~\cite{Jadach:2003bu} and the same range of $x\in(0.001,1)$.
In the upper plot of
fig.~\ref{fig:bremsCMC_glue} we show the gluon distribution evolved up 
to scale $Q=1$ TeV
(lower curve) due to pure gluon bremsstrahlung, obtained using our new CMC 
and the Markovian {\tt EvolMC}.
Results are indistinguishable; we therefore plot their ratio in the lower plot
of this figure.
The results of the two programs agree perfectly, within the statistical errors,
which are of order $0.5\%$ at low $x$.
MC results are for statistics of a few hundred millions of MC events.
A higher-statistics comparison will be presented in the following section.
We also include, in the upper plot of fig.~\ref{fig:bremsCMC_glue}, the
result of the evolution (from the Markovian MC)
due to all transitions, not only bremsstrahlung.
As we see, the complete result differs
from the pure bremsstrahlung one by a factor of almost 2, 
hence the inclusion of the
transitions of gluon into quark and back is very important.

In fig.~\ref{fig:bremsCMC_quark} we present 
numerical results of the analogous comparisons for the quark singlet
$Q=q+\bar{q}$. 
Again, pure bremsstrahlung results of the new CMC agree very well with these
of the Markovian {\tt EvolMC}, to within of the statistical error,
which is about 0.5\% in most of the $x$ range.
Contrary to the previous case 
the curve for full evolution coincides with that
for pure bremsstrahlung. This is easy to explain as a result of
the suppression of the gluon distribution at high $x$, resulting
in the smallness of the $Q\from G$ contribution.

On the technical side, let us remark that there are two methods of obtaining
pure bremsstrahlung contributions from the Markovian MC.
One may simulate full DGLAP evolution, including $Q\leftrightarrow G$ transitions
and select events (evolution histories) in which only pure 
bremsstrahlung occurs.
The other method is to suppress kernels for $Q\leftrightarrow G$ transitions
completely.
In the present version of the Markovian {\tt EvolMC},
both methods are available, and
both give identical results.
In the present work we mainly use the first method 
of selecting evolution histories
out of complete evolution.

\section{The CMC with the flavour transitions -- full DGLAP}
\subsection{General discussion}
Before getting into details, let us describe the essential ingredients
of our CMC algorithm for full LL DGLAP, with an arbitrary number of 
flavour-changing transitions $G\leftrightarrow Q$, where $Q=q,\bar{q}$.
The first ingredient is the observation, made in ref.~\cite{Jadach:2003bu},
that the average number $\langle n \rangle$ 
of $G\leftrightarrow Q$ transitions is
much lower than the average number 
$\langle N\rangle$ of $G\to G$ or $Q\to Q$ ones,
that is of the gluon bremsstrahlung emissions.
In fact $\langle n \rangle \simeq 1$ for 
the evolution from $Q_0=1$ GeV to $Q=1$ TeV,
while $\langle N\rangle \sim 20$ (for $\veps=10^{-4}$).
This suggests quite strongly that we should consider the evolution process
(emission chain) as a two-level process: sublevel of the pure bremsstrahlung
and superlevel of the flavour transitions.
Let us call it ``hierarchical'' organization of the emission chain.
The hope is that the small number of superlevel transitions
can be modelled, for example by a general-purpose MC tool such as {\tt FOAM},
thanks to a relatively small dimensionality of the problem.
The assumption is that pure bremsstrahlung segments can be treated separately
and efficiently.

As discussed in ref.~\cite{raport04-06},
for the treatment of the energy constraint $\delta$-function,
there are at least two options.
We may deal with it independently and separately for each of the two levels;
this is called  type I solution in ref.~\cite{raport04-06}.
In the other CMC solution,  nicknamed CMC type II in ref.~\cite{raport04-06},
the energy constraint is implemented globally,
for both levels at once, using the assumption (corrected later on by the MC weight)
that $D_k(t_0,x)\sim x^{\eta_k-1}$.
The latter solution is described and implemented in ref.~\cite{raport04-06},
up to bremsstrahlung level, with the explicit
algebraic layout for the full DGLAP.

In the following we shall walk along the path of CMC class I solution, 
the superlevel
implementation employing the general-purpose MC tool {\tt FOAM}
and the sublevel being implemented exactly as in the previous section.
Both levels feature the energy constraint, that is the total loss due
to multiple emissions/transitions is predefined and in the MC it is generated
as one of the first variables.
The same is true for the total number $n$ of flavour transitions and
the number of gluon emissions $n_i$,
in every $i$-th pure bremsstrahlung segment of the
emission chain.

\subsection{Hierarchical organization of the emission/evolution chain}
As we have argued in the above discussion, the two-level ``hierarchical''
reorganization of the emission/evolution chain is mandatory for any
reasonable CMC scenario.
In algebraic language,
the transition to hierarchical organization means that the sums over 
the flavour indices in the iterative solution of the evolution equations 
of eq.~(\ref{eq:Iter5})
are reorganized in such a way that all adjacent gluon emission vertices
are lumped together into the distributions/integrals $\Deu_{kk}(\tau,\tau_0;x)$
of the previous section.
The remaining integrals and sums will belong to the superlevel.
The formal derivation of the resulting hierarchical iterative solution
of the evolution equation is presented separately in 
ref.~\cite{raport04-09}.
In the following we shall present only the final result, discuss its
structure and apply it to the CMC algorithm.

\begin{figure}[!ht]
  \centering
  {\epsfig{file=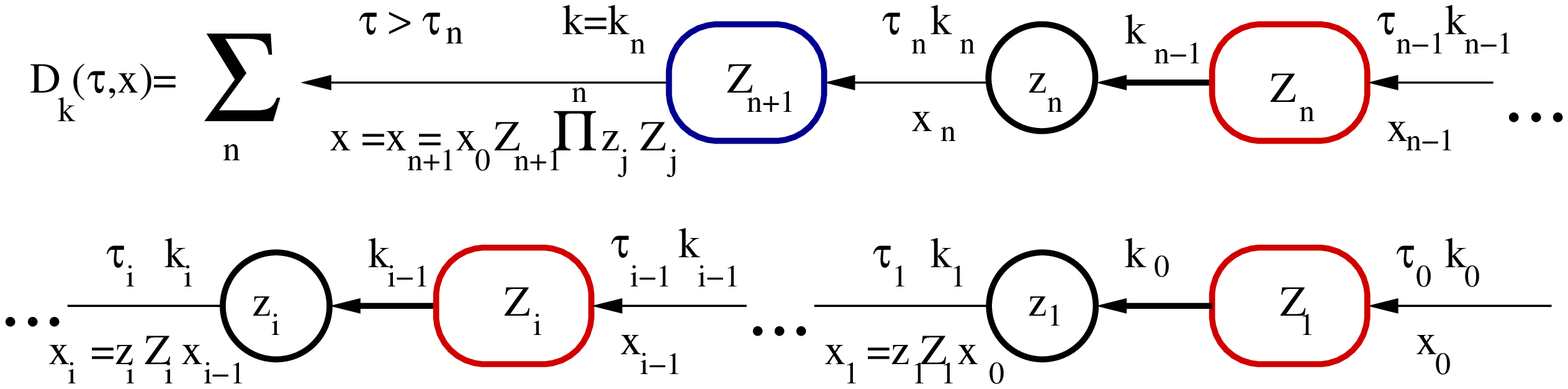,width=140mm}}
  \caption{\sf
    The scheme of kinematics and flavour indices in the hierarchical Markovian.
    }
  \label{fig:hierZ}
\end{figure}

The iterative solution of the DGLAP equation reorganized 
in the hierarchical form reads as follows:
\begin{equation}
  \label{eq:hierarchy}
  \begin{split}
  D_k&(\tau,x)=
     \int dZ\; dx_0\; 
     e^{-(\tau-\tau_0)R_k'}
     d_k(\tau,Z | \tau_{0})\;
     D_k(\tau_0,x_0)\delta_{x=Z x_0}+
\\&
  +\sum_{n=1}^\infty \;
   \sum_{{k_{n-1}\dots,k_{1},k_{0}}\atop %
         {k_{j}\neq k_{j-1}, j=1,\dots,n}}
   \bigg[ \prod_{j=1}^n \int\limits_{\tau_0}^\tau d\tau_j\; 
           \theta_{\tau_j>\tau_{j-1}} \bigg]\;
   \int\limits_0^1 dZ_{n+1}\;
   \bigg[ \prod_{i=1}^n \int\limits_0^1 dz_i\; 
                        \int\limits_0^1 dZ_i \bigg]\;
   \int\limits_0^1 dx_0\;
\\&~~~~~~~~~~~~~~~~~~\times
   e^{-(\tau-\tau_n)R_k'}\;\;
   d_{k}(\tau,Z_{n+1} | \tau_{n})\;
\\&~~~~~~~~~~~~~~~~~~\times
    \bigg[ \prod_{i=1}^n
     \Peu_{k_ik_{i-1}}^\Theta (z_i)\;
     e^{-(\tau_i-\tau_{i-1})R_{k_{i-1}}'}\;
     d_{k_{i-1}}(\tau_i,Z_i | \tau_{i-1}) \bigg]\;
\\&~~~~~~~~~~~~~~~~~~\times
   D_{k_0}(\tau_0,x_0) 
   \delta\bigg(x- x_0 \prod_{i=1}^n z_i \prod_{i=1}^{n+1} Z_i \bigg),
\\& 
d_k(\tau,Z|\tau_0)=
  Z^{-1}e^{-(\tau-\tau_0)R_{kk}} 
  \bigg\{ \delta_{Z=1}+
  \sum_{n=1}^\infty
   \prod_{i=1}^n 
     \int\limits_{\tau_0}^{\tau} d\tau_i\;\theta_{\tau_i>\tau_{i-1}}
     \int\limits_0^1 dz_i\;
	z_i\Peu_{kk}^\Theta (z_i) 
      \delta_{Z=\prod_{i=1}^n z_i} \bigg\},
\end{split}
\end{equation}
where we denote $k\equiv k_n$ and the virtual form factors are defined as
follows:
\begin{equation}
\label{eq:formfaktory}
\begin{split}
 &   R_k \equiv \Peu_{kk}^\delta(\veps) =
       \sum_j \int\limits_0^{1} dz\;
        z\Peu^{\Theta}_{jk}(z),\qquad
    \Phi_k(\tau,\tau_0)=(\tau-\tau_0)R_k, 
\\&
       R_{jk} = \int\limits_0^{1} dz\;
        z\Peu^{\Theta}_{jk}(z), \quad
    R_k' =
       \sum_{j\neq k} R_{jk} = R_k -R_{kk}.
\end{split}
\end{equation}
The $d_k$-function obeys the
normalization condition $\int_0^1 dZ \; Zd_k(\tau,Z|\tau_0) \equiv 1$.
It is easy to check that $d_k$ is
related to the pure brems\-strahlug distribution
$\Deu_{kk}$ worked out in section~\ref{sec:cmc-brems}, 
eq.~(\ref{eq:iterative}),
in the following way:
\begin{equation}
\begin{split}
&
  \Deu_{kk}(\tau,\tau_0;Z)=e^{-(\tau-\tau_0)R'_k} d_k(\tau,Z|\tau_0),
\\&
  R_k = b_k \ln(1/\veps) -a_k.
\end{split}
\end{equation}
As is also schematically  shown in fig.~\ref{fig:hierZ},
the integrand of eq.~(\ref{eq:hierarchy}) consists of a product of the
superlevel transition probabilities, each of them consisting of the
flavour-changing 
kernel and the pure bremsstrahlung function.
The whole emission chain is terminated with yet another
bremsstrahlung function.

Let us add the following interesting side remark:
We could interpret the above formula
as the Markovian process,
each step being an entire (pure bremsstrahlung) Markovian process
of its own. Of course,
it would be possible to implement such a hierarchical two-level Markovian
scenario as the MC simulation of the unconstrained QCD evolution%
\footnote{We tried to check in the literature whether this 
possibility was noticed
  by the authors of the classic Markovian MCs, but we could find no
  explicit reference to such a scenario.};
however, in such a case, contrary to the CMC case,
there is no convincing reason to invest an extra effort
to implement in practice such a solution.

\subsection{The CMC solution type I}
Exploiting the two-level hierarchical solution of eq.~(\ref{eq:hierarchy}),
we shall work out in the following the CMC solution of type I,
concentrating on the flavour-changing superlevel, 
because the pure bremsstrahlung level
is the same as described in section~\ref{sec:cmc-brems}, except that
it is repeated here not once but many times in a single 
evolution process.
As a first step, let us write eq.~(\ref{eq:hierarchy}) 
once again, in a more compact form, eliminating
$\int dx_0$ at the expense of the $\delta$-function:
\begin{equation}
\label{eq:hierarchy2}
\begin{split}
 &D_k(\tau,x)=
     \int dZ\;
     \Deu_{kk}(\tau,\tau_{0};Z)\;
     \frac{\bar{x}_0}{x} D_k(\tau_0,\bar{x}_0)\; \theta_{x<Z}+
\\&~~~
  +\sum_{n=1}^\infty \;
   \sum_{{k_{n-1}\dots,k_{1},k_{0}}\atop %
         {k_{j}\neq k_{j-1}, j=1,\dots,n}}
   \bigg[ \prod_{j=1}^n \int\limits_{\tau_0}^\tau d\tau_j\; 
          \theta_{\tau_j>\tau_{j-1}} \bigg]\;
   \int\limits_0^1 dZ_{n+1}\;
   \Deu_{kk}(\tau,\tau_{n};Z_{n+1})\;
\\&~~~~~~\times
    \bigg[ \prod_{i=1}^n
     \int\limits_0^1 dz_i\;
     \Peu_{k_ik_{i-1}}^\Theta (z_i)\;
     \int\limits_0^1 dZ_i\;
     \Deu_{k_{i-1}k_{i-1}}(\tau_i,\tau_{i-1};Z_i) \bigg]\;
\\&~~~~~~\times
   \frac{\bar{x}_0}{x} D_{k_0}(\tau_0,\bar{x}_0)\;
   \theta_{x < Z_{n+1}\prod_{i=1}^n z_iZ_i},
\\& \bar{x}_0 \equiv x\bigg( Z_{n+1}\prod_{i=1}^n z_iZ_i\bigg)^{-1}.
  \end{split}
\end{equation}
Next, we work out the explicit kinematic limits
in terms of $x$-variables defined as follows:
\begin{equation}
  x_i\equiv \bigg(\prod_{j=1}^i z_jZ_j\bigg) \bar{x}_0=
           x\bigg(Z_{n+1}\prod_{j=i+1}^n z_jZ_j\bigg)^{-1},
      \; i=0,1,2,..,n,\quad
  x_{n+1}\equiv x,\quad z_{n+1}\equiv 1.
\end{equation}
In this way, we obtain
\begin{equation}
\label{eq:hierarchy3}
\begin{split}
 &D_k(\tau,x)=
     x^{-1}\int_x^1 dZ_1\;
     \Deu_{kk}(\tau,\tau_{0};Z_1)\;
     \bar x_0 D_k(\tau_0,\bar x_0)+
\\&~~~
  +x^{-1}\sum_{n=1}^\infty \;
   \sum_{{k_{n-1}\dots,k_{1},k_{0}}\atop %
         {k_{j}\neq k_{j-1}, j=1,\dots,n}}
   \bigg[ \prod_{j=1}^n \int\limits_{\tau_0}^\tau d\tau_j\; 
          \theta_{\tau_j>\tau_{j-1}} \bigg]\;
   \int\limits_x^1 dZ_{n+1}\;
   \Deu_{kk}(\tau,\tau_{n};Z_{n+1})\;
\\&~~~~~~\times
   \bigg[  \prod_{i=1}^n
     \int\limits_{x_{i}}^1 dz_i\;
     \Peu_{k_ik_{i-1}}^\Theta (z_i)\;
     \int\limits_{x_{i}/z_i}^1 dZ_i\;
     \Deu_{k_{i-1}k_{i-1}}(\tau_i,\tau_{i-1};Z_i) \bigg]\;
   \bar x_0 D_{k_0}(\tau_0,\bar x_0).\;
  \end{split}
\end{equation}
The functions  $\Deu_{k_{i-1}k_{i-1}}(\tau_i,\tau_{i-1};Z_i)$ are
the multidimensional integrals of their own, 
described in section~\ref{sec:cmc-brems}, which in the Monte Carlo
are implemented as a separate module providing pure brems\-strahlung 
subevents with the weight $w^\#_{k_{i-1}}$.
In the first stage of the MC, neglecting $w=\prod_i w^\#_{k_{i-1}}$, 
i.e.~replacing $\Deu_{kk}$ by $\Deu'_{kk}$ of eq.~(\ref{deuprim}),
we have to generate
$3n+1$ {\em continuous variables} explicitly present 
in eq.~(\ref{eq:hierarchy3}) according to the integrand
\begin{equation}
\label{eq:hierarchy4}
\begin{split}
 &D_k(\tau,x)=
     x^{-1}
    \int\limits_0^{(1-x)^{b_k(\tau-\tau_{0})}} dU_{1}\;
    Z(U_{1})^{\omega_k-2} e^{a_{k}(\tau-\tau_{0})}\;
     \bar x_0 D_k(\tau_0,\bar x_0)+
\\&~
  +x^{-1}\sum_{n=1}^\infty \;
   \sum_{{k_{n-1}\dots,k_{1},k_{0}}\atop %
         {k_{j}\neq k_{j-1}, j=1,\dots,n}}
   \bigg[ \prod_{j=1}^n \int\limits_{\tau_0}^\tau d\tau_j\;
          \theta_{\tau_j>\tau_{j-1}} \bigg]\;
    \int\limits_0^{(1-x)^{b_k(\tau-\tau_{n})}} dU_{n+1}\;
    Z(U_{n+1})^{\omega_k-2} e^{a_{k}(\tau-\tau_{n})}\;
\\&~~~~~~~\times
  \bigg[ \prod_{i=1}^n
    \int\limits_{x_{i}}^1 dz_i\;
    \Peu_{k_ik_{i-1}}^\Theta (z_i)\;
    \int\limits_0^{(1-x_{i}/z_i)^{b_{k_{i-1}}(\tau_i-\tau_{i-1})}} dU_i\;
    Z(U_i)^{\omega_{k_{i-1}}-2} e^{a_{k_{i-1}}(\tau_i-\tau_{i-1})} \bigg]\;
\\&~~~~~~~\times
   \bar x_0 D_{k_0}(\tau_0,\bar x_0),
  \end{split}
\end{equation}
where $U_i=\exp(\Reu(Z_i))$ and
\begin{equation}
 \bar{x}_0 \equiv x\bigg( Z(U_{n+1})\prod_{i=1}^n z_iZ(U_i)\bigg)^{-1}.
\end{equation}
The first term ($n=0$) in the above sum is identical to the one
discussed in section~\ref{sec:cmc-brems}.
The second term for $n=1$ is the new and non-trivial one, representing
one $q\to G$ or $G\to q$ transition accompanied by 
the two segments of the pure brems\-strahlung.
It reads as follows:
\begin{equation}
\begin{split}
 &D_k(\tau,x)|_{n=1}=
  x^{-1} \sum_{{k_{0}}\atop %
               {k\neq k_{0}}}
   \int\limits_{\tau_0}^\tau d\tau_1\;
    \int\limits_0^{(1-x)^{b_k(\tau-\tau_{1})}} dU_{2}\;
    Z(U_{2})^{\omega_k-2} e^{a_{k}(\tau-\tau_{1})}\;
\\&~~~\times
      \int\limits_{x_{1}}^1 dz_1\;
      \Peu_{k k_{0}}^\Theta (z_1)\;
      \int\limits_0^{(1-x_{1}/z_1)^{b_{k_{0}}(\tau_1-\tau_{0})}} dU_1\;
      Z(U_1)^{\omega_{k_{0}}-2} e^{a_{k_{0}}(\tau_1-\tau_{0})}\;\;
        \bar x_0 D_{k_0}(\tau_0,\bar x_0),
  \end{split}
\end{equation}
where $x_1\equiv x/Z_2$, $\bar x_0\equiv x/(Z_2Z_1z_1)$ and $k_1\equiv k$.

In addition to continuous variables, we have to solve the problem of generating
efficiently all {\em discrete} variables $k_i, i=0,1,\dots, n-1$.
For gluon and $2n_f$ quarks and antiquarks 
in the contribution with $n$ flavour transitions,
we have in eqs.~(\ref{eq:hierarchy2})--(\ref{eq:hierarchy4})
up to $(2n_f+1)^n$ terms.
This might be a serious problem
in the case of implementing 
all these component integrals one by one
in the general-purpose MC tool such as {\tt Foam},
taking for example $n\leq n_{\max}=4$.
Luckily, 
thanks to symmetries valid in the  LL approximation,
we are able to get this problem under control, at least for
massless {\it identical quarks}. 
This methodology is not the same as the traditional splitting of
parton distributions into singlet and non-singlet parts,
although it exploits the same properties of the kernels.

The essence of the solution of the above problem can be demonstrated
in a transparent way for just one type of  quark and antiquark.
The extension to the case of $n_f$ identical massless quarks is not difficult
and will be done later on.
Let us analyse the flavour sum
$\sum_{k_{n-1}k_{n-2}\dots k_1,k_0}$, with the condition 
$k_i\neq k_{i-1}, i=1,2,\dots, n$, in eq.~(\ref{eq:hierarchy4}).
We split the parton distribution of eq.~(\ref{eq:hierarchy4}) into
components with a well defined number of flavour transitions
$D(x)=\sum_{n=0}^{n_{\max}} D^{(n)}$, and we shall discuss the
components $D^{(n)}$ one by one.
In the above and in the following we omit all bremsstrahlung 
contributions in $D^{(n)}$,
integrations over $z_i$, etc., in eq.~(\ref{eq:hierarchy4}) --
we keep track of only the flavour-changing kernels and their indices.

The easiest case is $n$=0, that is the case of the pure bremsstrahlung.
Here, the integrand contains just one term proportional to $D_k$,
modulo the bremsstrahlung part,
where $k$ is one of the three possible flavours $k=G,q,\bar{q}$.
In the distribution given to {\tt Foam} this contribution is treated separately
and is generated automatically by {\tt Foam} with the correct probability.

The first non-trivial case is that of $n=1$ and $k=k_1=G$.
Here, the sum under consideration is 
$D^{(1)}_G=\sum_{k_0} \Peu_{Gk_0} D_{k_0}$,
where $k_0\neq G$, hence $k_0=q,\bar{q}$.
Since $\Peu_{G q}=\Peu_{G \bar{q}}$ we may replace both of them by $\Peu_{G q}$
and pull them out of the sum.
We obtain
$D^{(1)}_G =\Peu_{G q} \sum_{k_0=q,\bar{q}} D_{k_0} = \Peu_{G q} D_{Q}$,
where we have introduced the inclusive (singlet) initial quark distribution
$D_{Q}=\sum_{k_0=q,\bar{q}} D_{k_0}$.

In the similar case $k=k_1=q$ for $n=1$, the sum under consideration is 
$D^{(1)}_q= \sum_{k_0} \Peu_{q k_0} D_{k_0}$.
In this case the only possible contribution is for $k_0=G$,
and the only remaining term is $D^{(1)}_q= \Peu_{q G} D_{G}$.
In the case of tagged antiquark, $k_0=\bar{q}$,
we obtain exactly the same contribution: $D^{(1)}_{\bar{q}}= \Peu_{q G} D_{G}$.

The general case of the $n>1$ transitions is quite similar to the $n=1$ case. 
In the case of $k=k_n=G$, in the sum
\begin{displaymath}
D^{(n)}_G=
\sum_{k_{n-1}k_{n-2}\dots k_1,k_0} 
   \Peu_{Gk_{n-1}}\dots \Peu_{k_{1}k_{0}} D_{k_0},
\end{displaymath}
we may repeat the previous reasoning we followed for $n=1$.
In the first step we reduce the summation over the last index 
$k_{n-1}=q,\bar{q}$ with the help of $\Peu_{Gq}=\Peu_{G\bar{q}}$ and
$\Peu_{qG}=\Peu_{\bar{q}G}$ obtaining:
\begin{displaymath}
D^{(n)}_G=
2\Peu_{Gq}\sum_{k_{n-2}\dots k_1,k_0} 
   \Peu_{qk_{n-2}}\dots \Peu_{k_{1}k_{0}} D_{k_0}.
\end{displaymath}
In the next step we get rid of the sum over $k_{n-2}$ 
\begin{displaymath}
D^{(n)}_G=
2\Peu_{Gq} \Peu_{qG} \sum_{k_{n-3}\dots k_1,k_0} 
   \Peu_{Gk_{n-3}}\dots \Peu_{k_{1}k_{0}} D_{k_0}.
\end{displaymath}
We continue with the elimination of the sums one by one,
obtaining for odd $n$:
\begin{displaymath}
D^{(n)}_G=
2^{(n-1)/2}\; \Peu_{Gq} \Peu_{qG} \Peu_{Gq} \dots \Peu_{Gq}\; D_{Q},
\end{displaymath}
while for even $n$ we obtain:
\begin{displaymath}
D^{(n)}_G=
2^{n/2}\; \Peu_{Gq} \Peu_{qG} \Peu_{Gq} \dots \Peu_{qG}\; D_{G}.
\end{displaymath}
As a result of the above reasoning,
the whole sum is reduced to just one term for each $n$.
(Every $n=0,1,\dots, n_{\max}$ is generated by {\tt Foam} separately.)
It is important to stress that, in the MC, where we are interested
in a fully exclusive history of the intermediate states in the emission chain,
we may easily undo the summation over equal contributions 
from quarks and antiquarks,
leading to factors $2^{(n-1)/2}$ or $2^{n/2}$,
and choose randomly with the probability $1/2$ between 
$k_j=q$ and $k_j=\bar{q}$ for every
intermediate non-gluon state (every other link in the emission tree).

The case of $k=k_n=q$ and $n>1$ can be analysed in an analogous way:
\begin{displaymath}
D^{(n)}_q=
\sum_{k_{n-1}k_{n-2}\dots k_1,k_0} 
   \Peu_{q k_{n-1}}\dots \Peu_{k_{1}k_{0}} D_{k_0}.
\end{displaymath}
Due to kernel properties and $k_{n}\neq k_{n-1}$,
the condition $D^{(n)}_q$ reduces in the first step to:
\begin{displaymath}
D^{(n)}_q=
\Peu_{qG}\sum_{k_{n-2}\dots k_1,k_0} 
   \Peu_{Gk_{n-2}}\dots \Peu_{k_{1}k_{0}} D_{k_0}.
\end{displaymath}
The final result is either
\begin{displaymath}
D^{(n)}_q=
2^{(n-1)/2}\; \Peu_{qG} \Peu_{Gq} \Peu_{qG} \dots \Peu_{Gq}\; D_{Q},
\hskip 1cm \hbox{for}~n~\hbox{odd}
\end{displaymath}
or
\begin{displaymath}
D^{(n)}_q=
2^{n/2}\; \Peu_{qG} \Peu_{Gq} \Peu_{qG} \dots \Peu_{qG}\; D_{G},
\hskip 1cm \hbox{for}~n~\hbox{even}.
\end{displaymath}

{\em Summarizing, for identical quarks, 
in the LL approximation, we may effectively get rid of summations
over all of the flavour indices!}
The case of $n_f$ identical quarks is quite analogous -- the only
difference is that we get $(2n_f)^{(n-1)/2}$ or $(2n_f)^{n/2}$ weight factors
in front of each single final term.
When restoring the type of the intermediate quark we choose randomly
with equal probability one of the $2n_f$ quarks and antiquarks.

The above collective treatment of the intermediate quarks/antiquarks
in the process of implementing the distributions of eq.~(\ref{eq:hierarchy4})
as integrands of {\tt Foam} is relatively easy
for the finite number of flavour transitions ($n=0,1,2,3,4$).
As we remember, the only case where we need to explicitly generate
an individual quark or antiquark type
$k_0$ according to $\sim\Peu_{k_0}(x_0) H_{k_0}$,
is the case of $n=0$, i.e. pure bremsstrahlung.
In all other cases, $n>0$,
we may treat quarks and antiquarks at the intermediate stage of the MC 
generation
collectively and randomly choose their individual type (index) later on,
with equal probabilities.

In the above explicit algebra, the difference
between the traditional split into singlet and non-singlet
components $D^{S}=D_q+D_{\bar{q}}$ and $D^{NS}=D_q-D_{\bar{q}}$ and 
our alternative split
into pure bremsstrahlung component $D_q^{(0)}$ or $D_{\bar{q}}^{(0)}$ 
and the rest $D_q^{(n>0)}=D_{\bar{q}}^{(n>0)}$ is manifest.
Both techniques exploit, of course,  the same symmetry 
properties of the kernels.

\begin{figure}[!ht]
  \centering
  {\epsfig{file=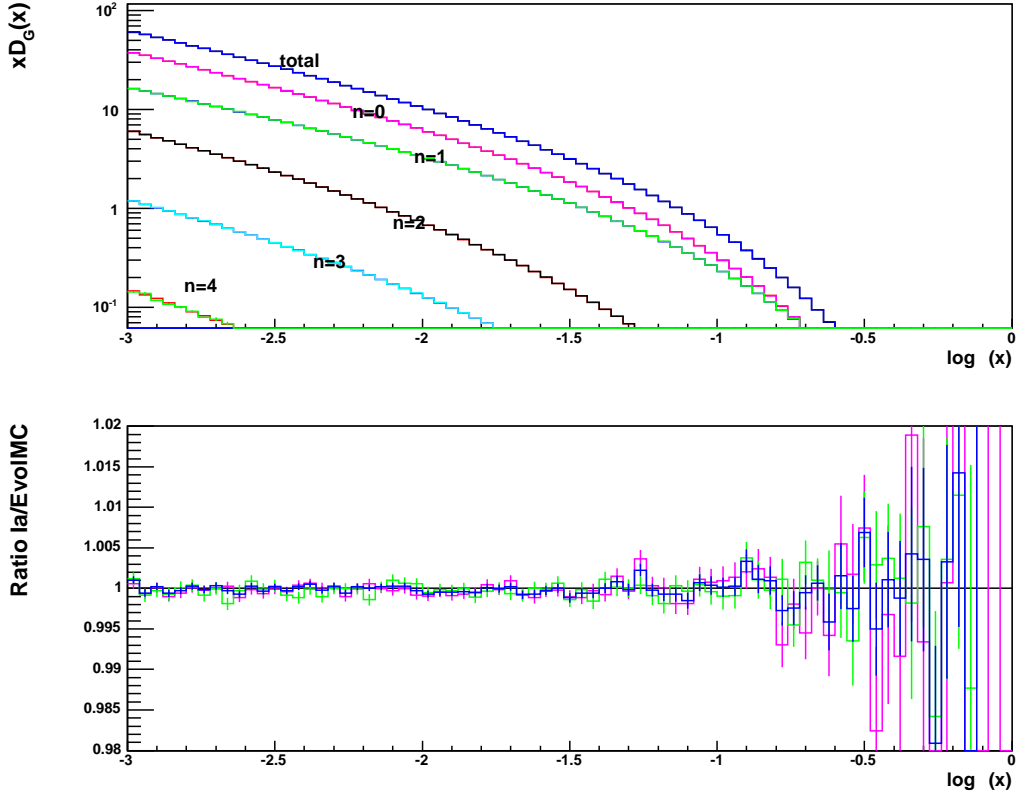, width=150mm}}
  \caption{\sf
    CMC versus Markovian MC for gluons;
    number of quark--gluon transitions $n=0,1,2,3,4$ and the total.
    The ratio in the lower plot is for $n=0,1$ and the total (blue).
    }
  \label{fig:CMCgluon}
\end{figure}
\begin{figure}[!ht]
  \centering
  {\epsfig{file=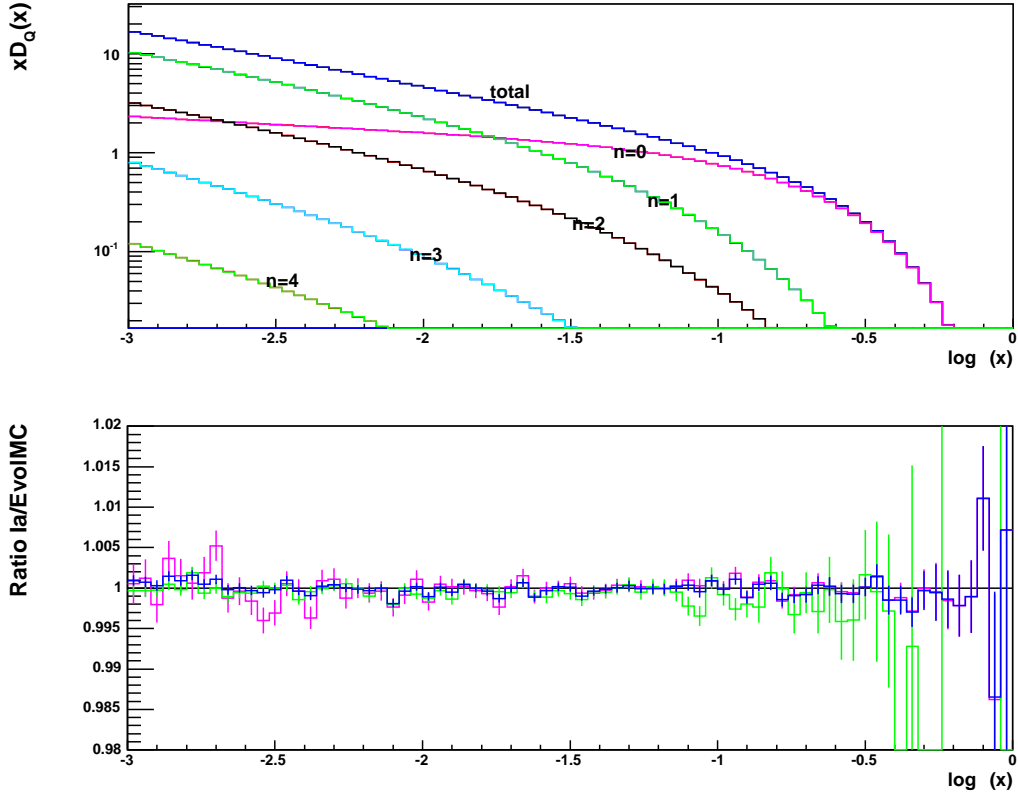, width=150mm}}
  \caption{\sf
    CMC versus Markovian MC for quarks; 
    number of quark-gluon transitions $n=0,1,2,3,4$ and the total.
    The ratio in the lower plot is for $n=0,1$ and the total (blue).
    }
  \label{fig:CMCquark}
\end{figure}
With all the above algebra at hand, we can formulate our
CMC algorithm in the general LL DGLAP case:
\begin{itemize}
\item
  Generate superlevel variables $n$, $k_i$, $\tau_i$ $Z_i$ and $z_i$
  using the {\tt Foam} general-purpose MC tool according to 
  eq.~(\ref{eq:hierarchy4}).
\item
  In the above we limit the number of flavour transitions 
  ($G\to Q$ and $Q\to G$)
  to $n=0,1,2,3,4$, aiming at a precision of $\sim 0.2\%$.
\item
  For each pure gluon brems\-strahlung segment
  defined by $Z_i$ and $(\tau_i,\tau_{i-1})$,
  $i=1,2,...,n+1$,
  gluon emission variables $(z^{(i)}_j,\tau^{(i)}_j),\; j=1,2,...,n^{(i)}$,
  are generated using the previously described dedicated CMC.
\item
  Weighted events are generated. They are optionally
  turned into weight-1 events using the conventional rejection method.
 \end{itemize}
The above algorithm is already implemented in the C++
programming language and tested using Markovian MC {\tt EvolMC}
of refs.~\cite{Jadach:2003bu} and \cite{Golec-Biernat:2006xw}; see next section.
In the program construction the central variable is
the number of quark--gluon transitions $n$.
We do not know the analytical expression for the distribution of this variable.
It is one of the variables managed by {\tt Foam}.
In the initialization phase, {\tt Foam} finds out the distribution of $n$ numerically
and then generates $n$ efficiently in the range $0 \leq n \leq n_{\max}$.
In the CMC program construction the $n_{\max}=0,1$ cases were programmed
first and the results were compared with {\tt EvolMC}.
The main programming exercise consist of the careful programming
of the integrand distribution for {\tt Foam}.
It was finally done for arbitrary $n_{\max}$.
The other part of the programming is related to joining pieces of several
pure bremsstrahlung segments into a single long emission chain
with an arbitrary number of quark--gluon transitions.
Object-oriented programming tools made this task easier.

\subsection{The CMC for DGLAP: numerical results}

The basic tests of the new CMC algorithm
are presented in figs.~\ref{fig:CMCgluon} and \ref{fig:CMCquark}.
As in section~\ref{sec:numtest0}, we examine results
of the DGLAP evolution from the energy scale $Q=1$ GeV to $Q=1$ TeV,
using as the starting point quark and gluon distributions
in proton at $Q=1$ GeV, exactly the same as in ref.~\cite{Jadach:2003bu}.
In fig.~\ref{fig:CMCgluon} we show distributions of gluon
at $Q=1$ TeV, while in fig.~\ref{fig:CMCquark} are
plotted the results for quark, $Q=q+\bar q$, at the same high scale $Q=1$ TeV.
In these two figures we compare gluon and quark distributions obtained
from the new CMC program and from the Markovian  
{\tt EvolMC}%
\footnote{This Markovian MC program 
    has been tested in refs.~\cite{Jadach:2003bu} and  \cite{Golec-Biernat:2006xw}
    against two non-MC programs {\tt QCDnum16}~\cite{qcdnum16}
    and {\tt APCheb33}~\cite{APCheb33}.
    Small systematic discrepancy between {\tt EvolMC} and {\tt QCDnum16} 
    for gluon distribution reported in ref.~\cite{Jadach:2003bu}
    was later eliminated and explained in ref.~\cite{Golec-Biernat:2006xw}.}
of ref.~\cite{Jadach:2003bu}.
The main numerical results from both programs,
marked as ``total'' in the upper plot of both figures,
are indistinguishable.
We therefore plot their ratio in the lower plot of both figures.
They agree perfectly well within the statistical error, 
in the entire range of $x$.
For $x<0.1$ the statistical error is below $0.1\%$.

The plots in figs.~\ref{fig:CMCgluon} and \ref{fig:CMCquark} contain, however,
more tests than that for the total normalization.
As already mentioned, in the process of constructing the CMC program
we have tested also each ``slice'' 
of the gluon and quark distribution for a given number
of quark--gluon transitions $n=0,1,2,3,4$, on the way from 1 GeV to 1 TeV.
For example, in fig.~\ref{fig:CMCgluon} we show separately the contributions
to the gluon distribution from the following evolution histories:\\
$n=0$: $G\to G$\\
$n=1$: $Q\to G$ and any number of gluon emissions out of $Q$ and $G$,\\
$n=2$: $G\to Q\to G$, etc.\\
$n=3$: $Q\to G\to Q\to G$, etc.\\
$n=4$: $G\to Q\to G\to Q\to G$, etc. 
           (``Total'' is the sum of $n=0,1,2,3,4$.)\\
They are shown in the upper plot of the figure one by one for 
the two programs compared, CMC and {\tt EvolMC}.
As before, the results are indistinguishable --
this is why we plot their ratios  in the lower plot of the figure.
The discrepancy is within the statistical error, as in the total
contribution.
In this plot the comparison of the two programs for the $n=0$ slice is a
repetition of the pure bremsstrahlung test from section~\ref{sec:numtest0},
but for much higher statistics.

Analogous slices for the quark distribution: \\
$n=0$: $Q\to Q$\\
$n=1$: $G\to Q$ and any number of gluon emissions out of $Q$ and $G$,\\
$n=2$: $G\to Q\to G\to Q$, etc.\\
$n=3$: $G\to Q\to G\to Q$, etc.\\
$n=4$: $Q\to G\to Q\to G\to Q$, etc. 
    (``Total'' is the sum of $n=0,1,2,3,4$)
are shown in fig.~\ref{fig:CMCquark}.
Again the results from the new CMC and Markovian {\tt EvolMC} agree
perfectly well.

\begin{figure}[!ht]
  \centering
  {\epsfig{file=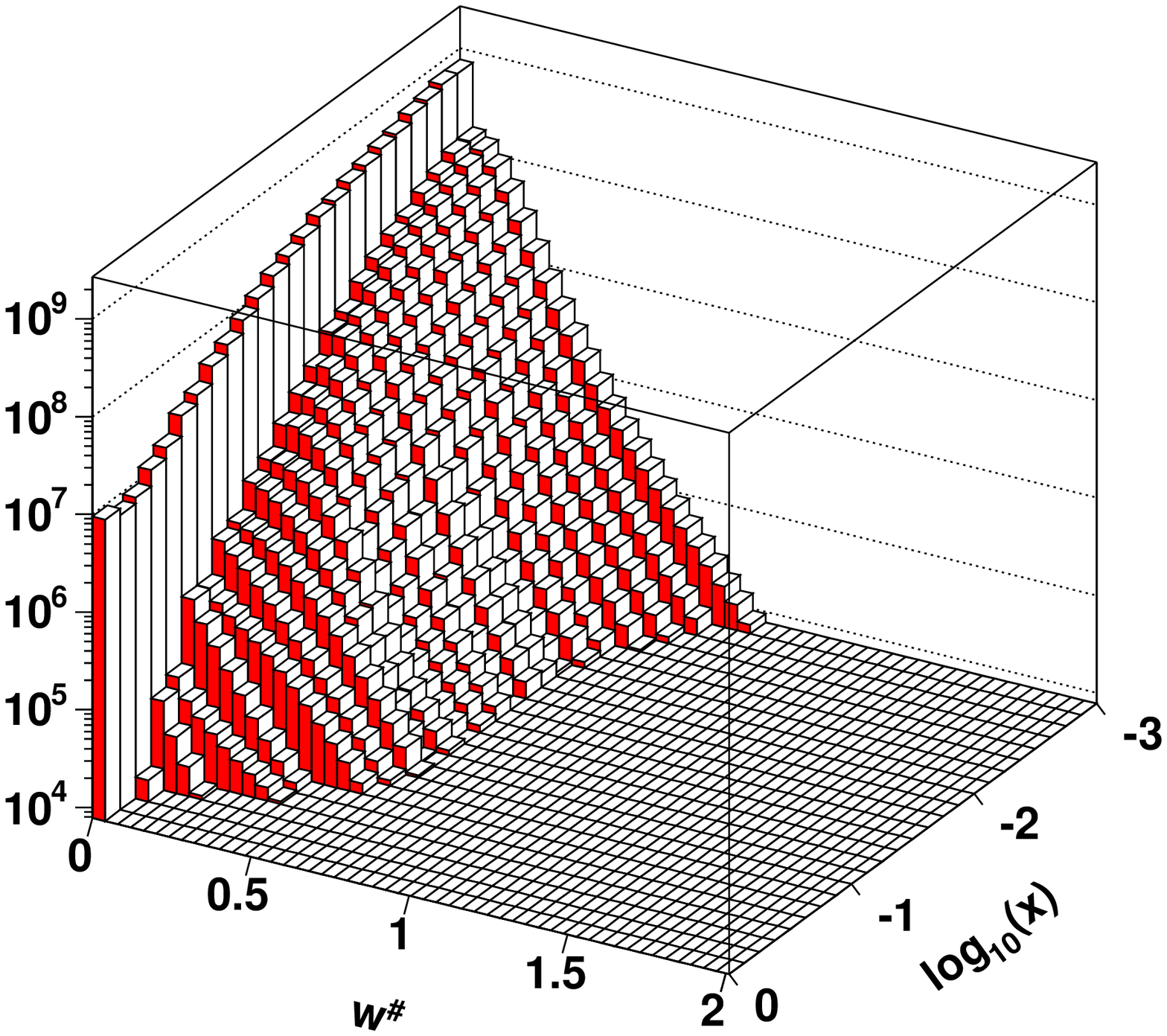, width=70mm}}
  {\epsfig{file=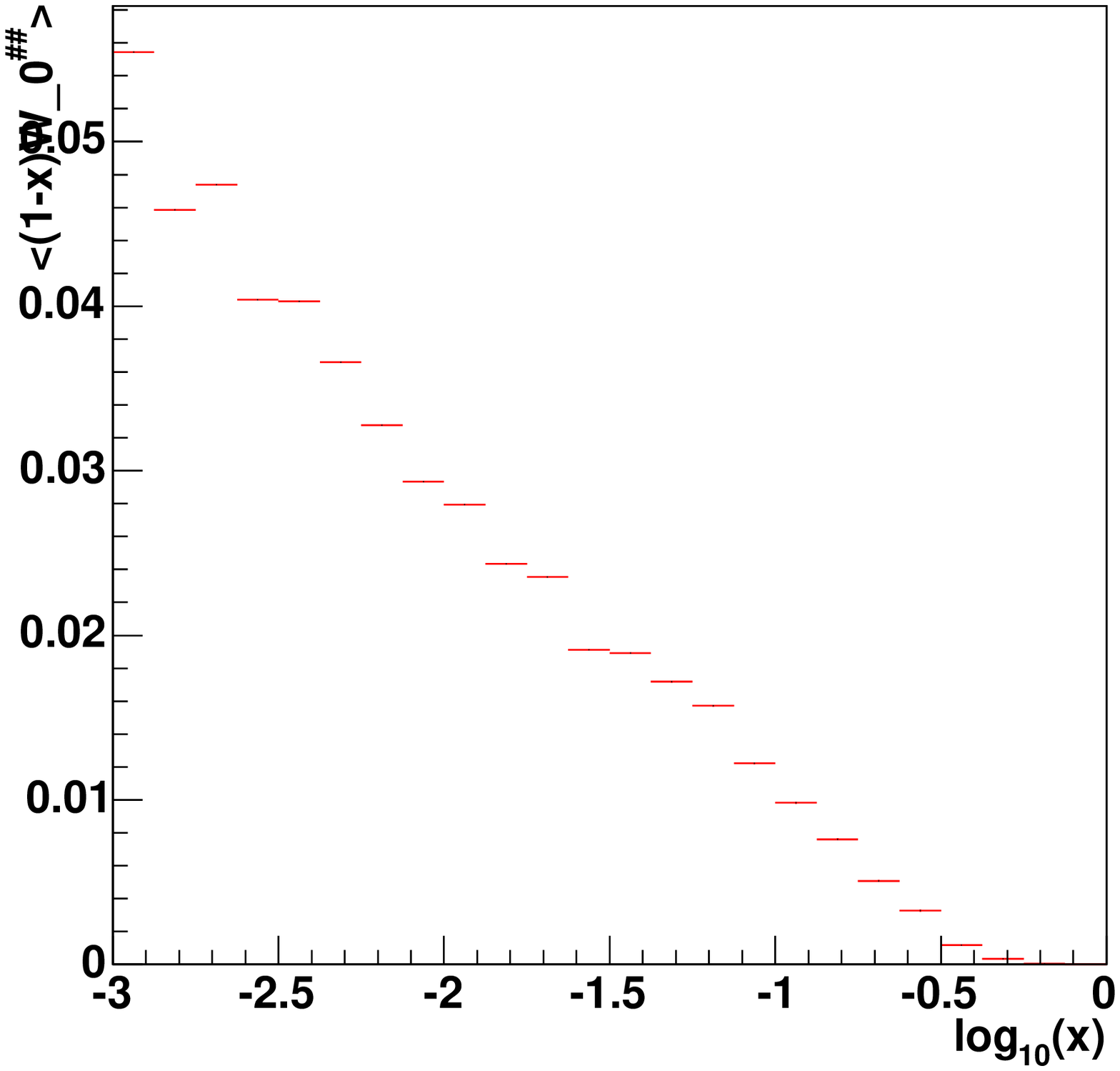, width=70mm}}
  {\epsfig{file=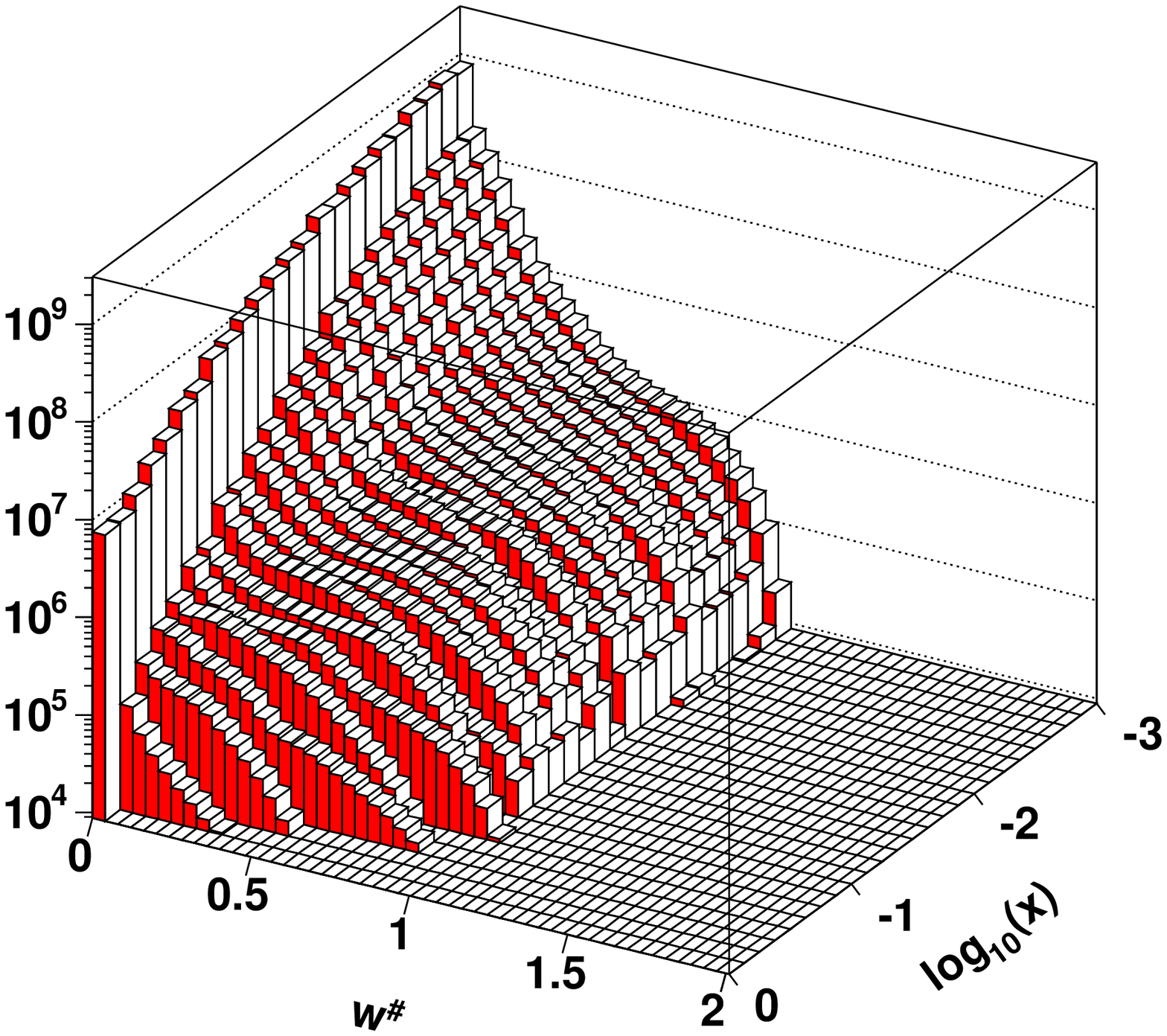, width=70mm}}
  {\epsfig{file=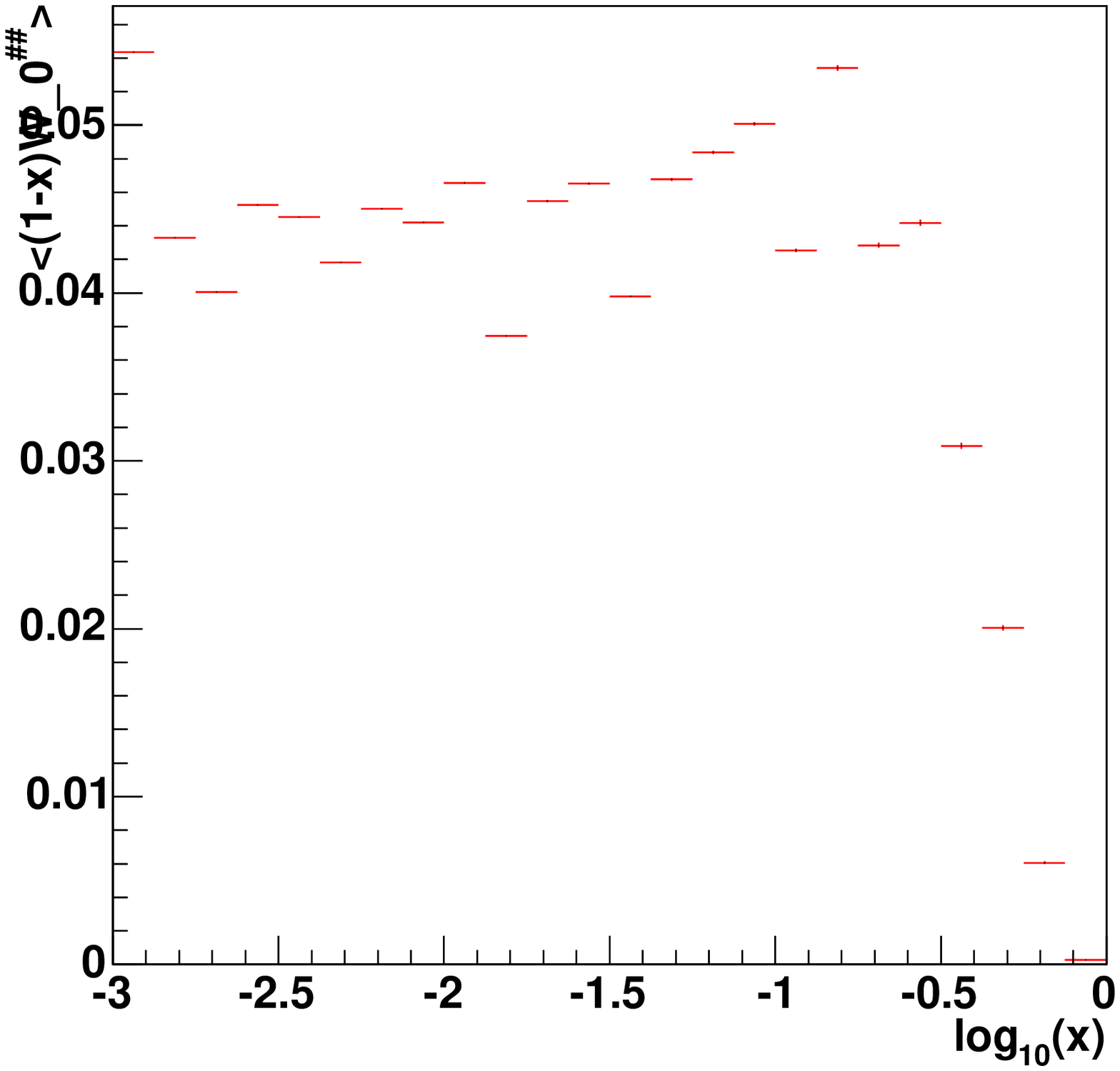, width=70mm}}
  \caption{\sf
    Weight distribution of the new CMC algorithm 
    for gluon (upper plots) and quark (lower plots)
    as a function of $x$.
    }
  \label{fig:CMCwtQ}
\end{figure}

In fig.~\ref{fig:CMCwtQ} we show the distributions of the  MC weight 
distribution
separately for the evolution yielding gluon and quark at $Q=1$ TeV.
This is an important technical test of the MC integration/simulation performed
with the help of  the general-purpose MC tool {\tt Foam}.
As we see, the resulting MC weight is well limited, $w\leq 1$,
and the average weight is about 0.08.
The efficiency is therefore quite good, substantially better than
that for method II.b of ref.~\cite{raport04-06}.
Note that for $n_{\max}=4$ the integrand of  {\tt Foam} is 15-dimensional,
but the modelling of this distribution is performed using only 2000 cells.
This clearly demonstrates the power and the usefulness of this tool.

\section{Conclusions and outlook}

In this work we have shown that there exists an {\em efficient} 
constrained Monte Carlo
algorithm for the initial-state radiation emission chain in QCD.
Its most likely application will be in the construction
of a new class of parton shower Monte Carlo event generators 
for the QCD initial-state radiation.

In this context it is worthwhile to mention that a similar CMC algorithm 
of class I
has been worked out~\cite{Jadach:2005rd} for the HERWIG-type QCD evolution, 
see ref.~\cite{Marchesini:1988cf},
i.e.\ for the $z$-dependent
$\alpha_S((1-z)Q)$ and $t$-dependent IR cut-off $\epsilon(t)$%
\footnote{This has been done for the case of pure gluonstrahlung combined with
   any number of quark-gluon transitions. 
   For more technical details see also http://home.cern.ch/jadach/public/desy\_mar05.pdf.}.

The present CMC solution is restricted to the LL kernels.
It will be interesting to extend it to the NLL case.
The preparatory step in this direction is already done.
In ref.~\cite{Golec-Biernat:2006xw}
the $\overline{MS}$ NLL kernels are implemented 
within the Markovian Monte Carlo program.
They can be ported to the non-Markovian CMC in the future, if necessary.

Let us stress that the CMC algorithm of this work is not the only one known.
For instance an alternative non-Markovian {\em  CMC algorithm class II}
exists; see refs.~\cite{zinnowitz04} and \cite{raport04-06}.
It is defined there for the full DGLAP, although it is implemented/tested
for the pure brems\-strahlung only.
It has worse MC efficiency than the algorithm presented here and leads to
higher dimensionality of the integrand managed by  {\tt FOAM}.

Another possible application of the presented CMC algorithm is
the MC modelling of the {\em unintegrated} parton distributions,
including these based on the CCFM-type evolution%
\footnote{ For example in the approach similar to that in
   the CASCADE Monte Carlo of ref.~\cite{Jung:2000hk}.},
for the purpose of MC simulating $W$ and $Z$ production at LHC.
In fact, the {\em unintegrated} parton distributions
$D_k(k_T,x)$ are already calculated~\cite{raport05-03} from 
the {\em one-loop type CCFM} model of ref.~\cite{Marchesini:1990zy},
in the framework of unconstrained Markovian MC.

\vspace{10mm}
\noindent
{\bf Acknowledgments}\\
We would like to thank W. P\l{}aczek and T. Sj\"ostrand for useful discussions.
We thank for their warm hospitality the CERN Particle Theory Group 
where part of this work was done.
We thank to ACK Cyfronet AGH Computer Center
for granting us access to their PC clusters
funded by the European Commission grant:
IST-2001-32243 
and the Polish State Committee for Scientific Research grants:
620/E-77/SPB/5PR UE/DZ 224/2002-2004 and
112/E-356/SPB/5PR UE/DZ 224/2002-2004

\appendix
\vfill\newpage
\section*{Appendix: QCD LL kernels}
We present here a table of the elements in the LL kernels 
($T_f=n_f T_R$), $Q=q+\bar{q}$
\begin{center}
\begin{tabular}{|r|r|r|r|r|r|r|}
\hline
$IK$& $A^{(0)}_{KK}$& $B^{(0)}_{KK}$& $C^{(0)}_{IK}$& $D^{(0)}_{IK}(z)$& $\hat{D}_{IK}(z)$
                                                           & $\int dz  D^{(0)}_{IK}(z)$ \\
\hline
$GG$ & $\frac{11}{6}C_A-\frac{2}{3}T_f$ & $2C_A$   
                           &   $2C_A$ & $2C_A(-2+z-z^2)$   & $0$ &  $-\frac{11}{3}C_A$\\
$QG$ &    $-$   &      $-$ &      $0$ & $2T_f(z^2+(1-z)^2)$& $2T_f$ &$\frac{4}{3}T_f$\\
$QQ$ & $\frac{3}{2}C_F$ & $2C_F$   
                           &      $0$ & $C_F(-1-z)$        & $0$ &  $-\frac{3}{2}C_F$\\
$GQ$ &    $-$   &      $-$ &   $2C_F$ & $C_F(-2+z)$        & $0$ &  $-\frac{3}{2}C_F$\\
\hline
\end{tabular}
\end{center}

\begin{equation}
  P_{ik}(z)= 
             \delta(1-z) \delta_{ik}      A_{kk}
            +\frac{1}{(1-z)_+}\delta_{ik} B_{kk}
	    +\frac{1}{z}                  C_{ik}
	                                 +D_{ik}(z).
\end{equation}
Temporary simplifications for the purpose of the MC generation are:
\begin{equation}
\begin{split}
  & zP^{\Theta}_{GG}(z) \to z\hat P^{\Theta}_{GG}(z)=
  zB_{GG}\theta_{1-z>\veps}
      \left(\frac{1}{1-z}+\frac{1}{z}\right)
   =  B_{GG}\frac{\theta_{1-z>\veps}}{1-z},
\\&
   zP^{\Theta}_{qq}(z) \to z\hat P^{\Theta}_{qq}(z)
   =  B_{qq}\frac{\theta_{1-z>\veps}}{1-z},
\\&
  z\Peu^{\Theta}_{kk}(t,z) 
  \to z\hat\Peu^{\Theta}_{kk}(t,z)
  = \frac{\alpha_S(t)}{\pi}\; z\hat P^{\Theta}_{kk}(z)
  = \frac{2B_{kk}}{\beta_0(t-t_\Lambda)}\;
         \frac{\theta_{1-z>\veps}}{1-z},
\\&
  \Peu^\delta_{kk}(t)=\frac{\alpha_S(t)}{\pi} 
     \bigg\{ B_{kk}\ln\frac{1}{\veps} -A_{kk} \bigg\}.
\end{split}
\end{equation}


\providecommand{\href}[2]{#2}

\end{document}